\documentclass[prd,aps,nofootinbib,eqsecnum,notitlepage,nobibnotes,superscriptaddress,preprintnumbers,floatfix,twocolumn]{revtex4-1}

\usepackage{subfig}
\usepackage[a4paper, left=20mm,right=20mm,top=20mm,bottom=20mm]{geometry}
\usepackage[titletoc,toc,title]{appendix}
\usepackage[normalem]{ulem}

\usepackage{tikz}
\usepackage{tabularx}

\usepackage{mathtools}
\usepackage{tensor}
\usepackage{amsmath, amsfonts,bm,amssymb}
\usepackage{derivative}
\usepackage{bm}

\usepackage{suffix}

\usepackage{amsmath, amsfonts,bm}
\usepackage{color}
\usepackage{graphicx}
\usepackage{subfig}
\definecolor{myred}{rgb}{0.9, 0.17, 0.31}

\usepackage[a4paper, left=20mm,right=20mm,top=20mm,bottom=20mm]{geometry}
\usepackage[colorlinks=true,hyperfootnotes=true, linkcolor=myred, citecolor=cyan, urlcolor=magenta]{hyperref}

\usepackage{array}
\usepackage{hhline}

\newcolumntype{Y}{>{\centering\arraybackslash}X}

\DeclareMathAlphabet{\mathup}{OT1}{\familydefault}{m}{n}

\def\dd{\mathrm{d}}

\newcommand{\be}{\begin{equation}} 
\newcommand{\ee}{\end{equation}}
\newcommand{\ba}{\begin{eqnarray}} 
\newcommand{\ea}{\end{eqnarray}}

\begin{document}

\title{Scaling solutions in three-form cosmology}

\author{Vitor da Fonseca}
\email{fc52156@alunos.ciencias.ulisboa.pt}
\affiliation{Instituto de Astrof\'isica e Ci\^encias do Espa\c{c}o,\\ 
Faculdade de Ci\^encias da Universidade de Lisboa,  \\ Campo Grande, PT1749-016 
Lisboa, Portugal}
\author{Bruno J. Barros}
\affiliation{Instituto de Astrof\'isica e Ci\^encias do Espa\c{c}o,\\ 
Faculdade de Ci\^encias da Universidade de Lisboa,  \\ Campo Grande, PT1749-016 
Lisboa, Portugal}
\author{Tiago Barreiro}
\affiliation{Instituto de Astrof\'isica e Ci\^encias do Espa\c{c}o,\\ 
Faculdade de Ci\^encias da Universidade de Lisboa,  \\ Campo Grande, PT1749-016 
Lisboa, Portugal}
\affiliation{ECEO, Universidade Lus\'ofona, Campo Grande, 376,  PT1749-024 Lisboa, Portugal}
\author{Nelson J. Nunes}
\affiliation{Instituto de Astrof\'isica e Ci\^encias do Espa\c{c}o,\\ 
Faculdade de Ci\^encias da Universidade de Lisboa,  \\ Campo Grande, PT1749-016 
Lisboa, Portugal}

\date{\today}

\begin{abstract}
A hybrid three-form model of dark energy is developed in order to identify scaling solutions, a long-sought feature in three-form cosmology. Exploiting Hodge dualities, the theory is formulated in terms of two scalar functions that are associated with the conjugate momentum, and the three-form dual vector in an isotropic background. The resulting Lagrangian yields a stable scaling attractor where the three-form energy density tracks the dominant background fluid. A dynamical mechanism is also identified that naturally drives the system out of this regime toward a late-time accelerated phase distinguishable from a cosmological constant.
This constitutes the first realization of scaling behavior within a three-form dark energy framework.
\end{abstract}

\maketitle
\section{Introduction}
\label{sec:intro}

Following the discovery of the universe's late-time acceleration~\cite{SupernovaCosmologyProject:1998vns,SupernovaSearchTeam:1998fmf,Caldwell:1997ii}, a dark energy component was postulated. Within the minimal and successful standard model of cosmology, this component corresponds to a cosmological constant. However, the long-standing cosmological constant problem~\cite{Peebles:2002gy} has motivated the exploration of evolving dark energy models. This idea has gained momentum with recent Dark Energy Spectroscopic Instrument (DESI) measurements of baryon acoustic oscillations~\cite{DESI:2024mwx}. Combined with luminosity-distance measurements of distant supernovae~\cite{Rubin:2023jdq,Scolnic:2021amr,DES:2024jxu} and Cosmic Microwave Background observations~\cite{Planck:2018vyg,ACT:2023dou,ACT:2023kun},  they have revealed tensions within the standard $\Lambda$CDM framework, hinting at departures from a strictly constant dark energy component. The topic is currently under active discussion within the scientific community (see, e.g., Refs.~\cite{Cortes:2024lgw,Wolf:2024eph,Linder:2025zxb} and references therein).

Among possible scenarios, scaling solutions are particularly appealing, as they allow the dark energy density to evolve dynamically and to remain non-negligible in the early universe. In such configurations, the energy scale of the dark component can approach that of radiation or matter, thereby alleviating the coincidence problem. More importantly, these solutions often exhibit attractor behavior, being weakly sensitive to initial conditions and effectively providing a self-tuning mechanism. In this picture, the observed energy composition of the universe today can be viewed as the outcome of an attractor-like evolution. However, an additional process is required to enable an exit from the scaling regime so that dark energy eventually dominates at late times, in agreement with present observations.

Within the framework of quintessence models~\cite{Wetterich:1994bg,Zlatev:1998tr,Chiba:1999ka,dePutter:2007ny}, such scaling attractors have been extensively studied and successfully realized~\cite{Copeland:1997et,Steinhardt:1999nw,Barreiro:1999zs,Tsujikawa:2004dp,Tsujikawa:2006mw}. In these models, the scalar field typically evolves dynamically during the early universe, before its potential drives the transition to a late-time accelerated phase~\cite{Copeland:2006wr}. In contrast, realizing such scaling behavior within a three-form dark energy framework has so far proven elusive~\cite{Koivisto:2009fb,Koivisto:2009ew,Wongjun:2016tva,Wongjun:2017spo,daFonseca:2024boz}, 
despite the fact that three-form fields can drive accelerated cosmological dynamics relevant to both the early and late universe~\cite{Koivisto:2009ew,Morais:2016bev,Bouhmadi-Lopez:2016dzw}.
Although the theoretical foundations of three-form cosmology have been extensively developed~\cite{Koivisto:2009ew,Wongjun:2016tva,Morais:2017vlf,Bouhmadi-Lopez:2018lly,Bouhmadi-Lopez:2020wve,Bouhmadi-Lopez:2021zwt,Barros:2023nzr}, 
it is only recently that observational constraints, combining low- and high-redshift probes, have been derived for a three-form dark energy model~\cite{Bouhmadi-Lopez:2025lzm}. The latter
shows potential for addressing the Hubble tension \cite{CosmoVerseNetwork:2025alb}, yet without cosmological scaling solutions.

In this paper, we present a novel hybrid framework that unveils scaling solutions within a three-form dark energy model for the first time.
In this formulation, we construct a Lagrangian expressed in terms of two potentials, $U(\phi)$ and $V(\chi)$, where the dual $\phi$ is associated with the conjugate momentum that generalizes the field-strength tensor, and $\chi$ describes the three-form dual vector compatible with a isotropic background. We find the functional forms of $U$ and $V$ that yield scaling dynamics between dark energy, described by a three-form field, and the radiation and matter-dominated background throughout cosmic evolution.
The resulting scaling regimes exhibit attractor behavior.
While the framework can be recast as an equivalent k-essence model, it remains worthwhile to investigate the intrinsic dynamics of three-form theories themselves, as the dualization to scalar fields can sometimes become intractable~\cite{BeltranAlmeida:2018nin,Duff:1980qv,Koivisto:2009fb}.
Furthermore, by extending the dynamical system, we identify a family of solutions that naturally exit the scaling regime, leading to dark energy domination and driving the late-time accelerated expansion of the universe.

This manuscript is organized as follows. In Sec.~\ref{sec:1st3form}, we introduce a three-form first-order formalism, generalizing its kinetics term through a function of the conjugate momentum. We aim to determine the form of this function that achieves scaling behavior. To do so, we consider the Hodge dual formulation applied to an isotropic spacetime and specify its implementation within a Friedmann-Lema\^itre-Robertson-Walker (FLRW) metric. In Sec.~\ref{sec:scaling}, we derive the three-form Lagrangian that gives rise to scaling solutions, which include a trivial case equivalent to a quintessence field with an exponential potential. Section~\ref{sec:system} analyzes the attractor behavior of the general scaling solution through the study of the associated dynamical system. In Sec.~\ref{sec:exit}, we explore a mechanism that enables the system to exit the scaling regime and evolve toward dark energy domination while preserving the structure of the original Lagrangian. We then examine the cosmological implications of the proposed model. Sec.~\ref{sec:conclusions} concludes on the viability of the cosmology obtained. Appendix~\ref{app:eigenvalues} provides for more details on the analysis of the dynamical system. Appendix~\ref{app:k-Lagrangian} demonstrates the consistency between the three-form scaling solution and the class of k-essence models that also admit scaling behavior. 

\section{The hybrid formalism}
\label{sec:1st3form}

\subsection{First order formulation of a massive three-form effective field theory}

Let us start by considering a massive three-form field $A_{\mu\nu\rho}$ within a first order Lagrangian formulation,
\be\label{eq:action}
\mathcal{L}=\frac{1}{48}f(\Pi^2)+\frac{1}{6}\Pi^{\mu\nu\rho\sigma}\nabla_{[\mu}A_{\nu\rho\sigma]}-V(A^2)\,,
\ee
where $\Pi_{\mu\nu\rho\sigma}$ is an arbitrary four-form auxiliary field, and the notation for squaring a tensor indicates full contraction of its indices, i.e., ${A^2\equiv A^{\mu\nu\rho}A_{\mu\nu\rho}}$. The function $f(\Pi^2)$ generalizes the kinetic term, while ${V(A^2)\neq0}$ represents a self-interaction potential. For simplicity, we omit the explicit indices throughout this section.

To recover a second-order theory, the auxiliary field $\Pi$ can be integrated out by solving its algebraic equation of motion,
\be\label{eq:aux:1}
\Pi f'(\Pi^2)=-4\left[\nabla A\right]\,,
\ee
where a prime denotes differentiation with respect to the argument in parentheses.  Substituting this relation back into Eq.~\eqref{eq:action} yields a Lagrangian expressed purely in terms of the three-form field and its derivatives,
\be
\mathcal{L}(A,\nabla A)=\left[\frac{1}{48}\bigg(f(\Pi^2)-2\Pi^2f'(\Pi^2)\bigg)-V(A^2)\right]_{\Pi(\nabla A)}
\ee

We note that in the simple case ${f(\Pi^2)=-\Pi^2}$, Eq.~\eqref{eq:aux:1} identifies the conjugate momentum $\Pi$ with the field-strength tensor of the three-form, $\Pi=\dd A$. Substituting this relation back into Eq.~\eqref{eq:action} recovers the canonical massive three-form Lagrangian \cite{Germani:2009iq}, corresponding to the standard kinetic term.

Conversely, we may derive the dynamical equations for the three-form $A$,
\be\label{eq:aux:2}
\nabla\cdot\Pi = -12 AV'(A^2)\,,
\ee
which allows us to find the relation ${A(\nabla\cdot\Pi)}$ and thereby rewrite the original Lagrangian as an effective four-form theory:
\be
\mathcal{L}(\Pi,\nabla \Pi)=\left[\frac{1}{48}f(\Pi^2)+2A^2V'(A^2)-V(A^2)\right]_{A(\nabla\cdot \Pi)}\,,
\ee
where the mixed term in Eq.~\eqref{eq:action} has been integrated by parts.

\subsection{Hodge dualities take the stage}

We now introduce the Hodge duals to the fields $\Pi$ and $A$,
\ba
\left(\star\Pi\right)=\frac{1}{4!}\varepsilon_{\mu\nu\rho\sigma}\Pi^{\mu\nu\rho\sigma}=:\phi\,\,\Rightarrow\,\,\Pi_{\mu\nu\rho\sigma}=-\varepsilon_{\mu\nu\rho\sigma}\phi\,,\quad \\
\left(\star A\right)_\mu=\frac{1}{3!}\varepsilon_{\mu\nu\rho\sigma}A^{\nu\rho\sigma}=:B_\mu\,\,\Rightarrow\,\,A_{\nu\rho\sigma}=\varepsilon_{\mu\nu\rho\sigma}B^\mu\,,\quad \label{eq:dual2}
\ea
which are a scalar and a vector, $\phi$ and $B^\mu$, respectively. Here $\varepsilon_{\mu\nu\rho\sigma}$ is the generalized Levi-Civita symbol. These Hodge duals let us map high-rank antisymmetric tensors to more tractable quantities, simplifying the dynamical analysis. In four spacetime dimensions, the three-form $A$ behaves dynamically like a vector field $B^\mu$ while the four-form $\Pi$ behaves like a scalar field $\phi$. Thus, we can square the above relations to find,
\be
\Pi^2=-24\phi^2\quad\text{and}\quad A^2 = -6B^2\,.
\ee
This allows us to rewrite the initial Lagrangian, Eq.~\eqref{eq:action}, in a hybrid representation of a vector and a scalar,
\be\label{eq:action_dual}
\mathcal{L}=\frac{1}{48}f(-24\phi^2)-\phi\nabla\cdot B-V(-6B^2)\,.
\ee
Again, the three-form $A$ is now expressed through the (Hodge) 1-form $B$ and the auxiliary four-form $\Pi$ through the 0-form (scalar) $\phi$. Performing integration by parts in the middle term of Eq.~\eqref{eq:action_dual}, we can write
\be\label{eq:action_dual2}
\mathcal{L}=\frac{1}{48}f(-24\phi^2)+B\cdot\nabla\phi-V(-6B^2)\,.
\ee
Naturally, both Eq.~\eqref{eq:action_dual} and \eqref{eq:action_dual2} share the same equations of motion, which read:
\ba
\nabla\cdot B &=& -\phi f'(-24\phi^2)\,, \label{eqmov1} \\
\nabla\phi &=& -12BV'(-6B^2)\,. \label{eqmov2}
\ea
These are also obtainable by direct dualization of Eqs.~\eqref{eq:aux:1} and \eqref{eq:aux:2}.

The latter Lagrangian, Eq.~\eqref{eq:action_dual2}, gives the following components for the energy momentum tensor,
\ba\label{eq:EMtensor}
T_{\mu\nu} &=& -2\frac{\partial\mathcal{L}}{\partial g^{\mu\nu}}+g_{\mu\nu}\mathcal{L} \nonumber \\
&=& -2B_\mu\nabla_\nu\phi-12V'(-6B^2)B_\mu B_\nu + g_{\mu\nu}\mathcal{L}\,,\quad\quad\quad
\ea
where $g_{\mu\nu}$ is the spacetime metric, which is not yet particularized.
\subsection{Isotropic background}

Let us now adopt a parametrization compatible with a FLRW isotropic metric with signature ${(-+++)}$. Assuming spatial flatness, the line element is given by \cite{1922ZPhy...10..377F,1931MNRAS..91..483L,1936ApJ....83..257R},
\be
\dd s^2=-\dd t^2+a^2(t)\eta_{ij}\dd x^i\dd x^j\,,
\ee
for spatial indices $i,j$ and cosmic time $t$, where $a$ is the dimensionless scale factor that parametrizes the expansion of the universe, and $\eta_{ij}$ denotes the three-dimensional Euclidean spatial metric.

The three-form dual vector can be written with the aid of a scalar function $\chi(t)$, dependent only on cosmic time at background level \cite{Germani:2009iq}. Thus, its components are
\be\label{eq:Bflrw}
B^0=\chi(t)\quad\text{and}\quad B^i=0\,,
\ee
and the invariant $B^2$ reads
\be
B^2=B^\mu B_\mu=B^0B_0=-\chi^2\,,
\ee
in the isotropic background. Additionally, from Eq.~\eqref{eq:dual2}, the non-vanishing components of the three-form are
\be
A_{ijk}=\sqrt{-g}\,\epsilon_{ijk}\chi(t)=a^3(t)\chi(t)\,,
\ee
where $g$ is the metric determinant and $\epsilon_{ijk}$ the standard three-dimensional Levi-Civita. At linear level, this allows us to recast the Lagrangian of the hybrid formalism, Eq.~\eqref{eq:action_dual2}, as a function of two scalars, $\phi$ and $\chi$.

Accordingly, the stress-energy tensor in Eq.~\eqref{eq:EMtensor} with the parametrization in Eq.~\eqref{eq:Bflrw} reads
\ba\label{eq:EMtensor2}
T_{\mu\nu} =&-&2B_\mu\nabla_\nu\phi
-\frac{V_{,\chi}}{\chi}B_\mu B_\nu \nonumber\\
 &+& g_{\mu\nu}\Big[-U(\phi)-V(\chi)+B^\alpha\nabla_\alpha\phi\Big]\,,\quad
\ea
where we have defined $U$ as
\be\label{eq:U_definition}
U=-\frac{f}{48}\,,
\ee
recalling that the function $f$ generalizes the kinetic term of the massive three-form in the original Lagrangian \eqref{eq:action}. If the three-form is treated as a perfect fluid, the temporal and spatial components of the energy-momentum tensor, $T^0{}_0=-\rho_A$ and $T^i{}_j=p_A\delta^i{}_j$, respectively, define the energy density and pressure of the three-form in the FLRW metric as,
\ba
\label{eq:rho_T00}
\rho_A &=& U(\phi)+V(\chi)-\chi V_{,\chi}+\chi\dot{\phi}\,, \\
\label{eq:p_Tij}
p_A &=& -U(\phi)-V(\chi)+\chi\dot{\phi}\,,
\ea
where an overdot denotes differentiation with respect to cosmic time.

Owing to the first-order nature of the formalism, the dynamics of the scalars are governed by two first-order equations of motion, which are mathematically coupled. Rewriting the equations of motion Eqs.~\eqref{eqmov1} and \eqref{eqmov2} in the isotropic background gives
\begin{align}
\label{eq:dot_chi}
\dot{\chi}+3H\chi &= -U_{,\phi} \,, \\
\label{eq:dot_phi}
\dot{\phi} &=  V_{,\chi}\,.
\end{align}
Using Eq.~\eqref{eq:dot_phi}, the on-shell three-form energy density and pressure are
\ba
\label{eq:rho_A}
\rho_A &=& U(\phi)+V(\chi)\,, \\
\label{eq:p_A}
p_A &=& -U(\phi)-V(\chi)+\chi V_{,\chi}\,,
\ea
respectively. Since $U$ and $V$ play the role of self-interacting terms in the equations of motion Eqs.~\eqref{eq:dot_chi} and \eqref{eq:dot_phi}, we will simply refer to them as potentials.

Although $\phi$ and $\chi$ are not directly observable, they fully encode the dynamics of the original three-form field whose mixed Lagrangian reads,
\be
\label{eq:hybrid}
\mathcal{L}= -U(\phi)-V(\chi)+\chi V_{,\chi}\,.
\ee

While the conventional massive canonical three-form (with a standard kinetic term and typical self-interacting potentials) fails to produce scaling solutions ~\cite{Koivisto:2009fb,Koivisto:2009ew,daFonseca:2024boz}, we explore whether such cosmologies can emerge from the hybrid model~\eqref{eq:hybrid} within a general relativistic FLRW background. 

\section{Scaling solutions}
\label{sec:scaling}

\subsection{Setting the cosmological background}

We embed the effective field theory within standard Einstein gravity by adding an Einstein-Hilbert term to the total action. The resulting field equations yield the Friedmann constraint,
\be
\label{eq:friedmann}
3H^2 = V(\chi) + U(\phi)+\rho_\gamma\,,
\ee
in natural units, where $H=\dot{a}/a$ is the Hubble function. The energy density $\rho_\gamma$ represents that of an additional background fluid with a barotropic index $\gamma$ satisfying $0<\gamma<2$: ${\gamma=4/3}$ for radiation and ${\gamma=1}$ for pressureless matter. The continuity equation for this background component is then
\begin{equation}
    \rho_\gamma'=-3\gamma\rho_\gamma\,,
\end{equation}
where a prime denotes differentiation with respect to the number of $e$-folds, $N\equiv\ln a$. Integrating it gives
\begin{equation}
\label{eq:diluation_background}
    \rho_\gamma\propto e^{-3\gamma N}\,.
\end{equation}

As for the three-form fluid, the energy conservation reads
\begin{equation}
    \label{eq:conservation_A}
    \rho_A^\prime=-3\left(\rho_A+p_A\right)\,,\quad\quad \rho_A+p_A=\chi V_{,\chi}\,,
\end{equation}
according to Eqs.~\eqref{eq:rho_A} and \eqref{eq:p_A}.

Our goal is to identify scaling solutions in cosmological regimes where the universe can be approximately described by radiation or pressureless matter (baryons and dark matter), together with a dark energy component represented by the three-form field. In particular, we are interested in solutions for which the energy densities of the dominant background fluid and dark energy remain of the same order of magnitude over an extended period, thereby alleviating the cosmological coincidence problem \cite{Peebles:2002gy}.

Denoting the fractional energy densities of the three-form and the dominant background fluid by
\be
\Omega_A=\frac{\rho_A}{3H^2}\,,\quad\quad \Omega_\gamma=\frac{\rho_\gamma}{3H^2}\,,
\ee
respectively, a scaling regime is characterized by a constant, non-zero ratio $\Omega_A/\Omega_\gamma$~\cite{Wetterich:1987fm,Wands:1993zm}. Using Eqs.~\eqref{eq:rho_A} and \eqref{eq:diluation_background}, and assuming ${U\propto V}$ in the scaling, we may write
\begin{equation}
    \label{eq:rho_A_scaling}    
    U \propto V \propto 
\rho_A\propto\rho_\gamma\propto e^{-3\gamma N}\,,
\end{equation}
that is, the potentials dilute at the same rate
\be
\label{eq:same_rate}
\frac{V'}{V}=\frac{U'}{U}=-3\gamma\,,
\ee
during the cosmological expansion, which satisfies
\begin{equation}
\label{eq:H_scaling}
    \frac{H^\prime}{H}=-\frac{3\gamma}{2}\,.
\end{equation}
This is analogous to the constancy of the kinetic-to-potential energy ratio in the scaling regime of a canonical scalar field \cite{Piazza:2004df}.

With Eqs.~\eqref{eq:conservation_A} and \eqref{eq:rho_A_scaling}, we have
\begin{equation}
\label{eq:equiv}
    \rho_A+p_A=\chi V_{,\chi}\propto e^{-3\gamma N}\,,
\end{equation}
and therefore $\chi V_{,\chi}\propto e^{-3\gamma N}$ is the general condition that characterizes scaling solutions for the three-form Lagrangian \eqref{eq:hybrid} derived from first-order formulation.

For the record, we can introduce the adiabatic sound speed of the three-form background, defined by
\begin{equation}
    c_a^2=\frac{p_A^\prime}{\rho_A^\prime}\,.
\end{equation}
Since $\rho_A^\prime=-3(\rho_A+p_A)=-3\chi V_{,\chi}$, the sound speed reads
\begin{align}
    \label{eq:c2}
    c_a^2=-\frac{(\chi V_{,\chi})^\prime}{3\chi V_{,\chi}}-1\,.
\end{align}
Under the scaling condition in Eq.~\eqref{eq:equiv}, it reduces to
\begin{equation}
    c_a^2=\gamma-1\,,
\end{equation}
along a scaling trajectory, as expected.

\subsection{Finding the general scaling form for $V$ and $U$}

To find the form of the potential $V(\chi)$ that realizes a scaling solution from the general condition in Eq.~\eqref{eq:equiv}, we write
\begin{equation}
\label{eq:diff}
    \frac{d\ln{(\rho_A+p_A)}}{dN}=\frac{\partial\ln(\chi V_{,\chi})}{\partial\ln\chi}\frac{\chi^\prime}{\chi}=-3\gamma\,.
\end{equation}
Moreover, since Eq.~\eqref{eq:equiv} can be written as
\begin{equation}
    \frac{\chi}{\chi^\prime}V'\propto e^{-3\gamma N}\,,
\end{equation}
and ${V'\propto e^{-3\gamma N}}$ according to Eq.~\eqref{eq:rho_A_scaling}, $\chi^\prime/\chi$ is a constant that we call $\alpha\neq0$:
\begin{equation}
\label{eq:alpha}
\frac{\chi^\prime}{\chi}=\alpha\,.
\end{equation}
Substituting Eq.~\eqref{eq:alpha} in Eq.~\eqref{eq:diff}:
\begin{equation}
    \label{eq:solution_scaling}
    \frac{\partial\ln(\chi V_{,\chi})}{\partial\ln\chi}=-\frac{3\gamma}{\alpha}\,,
\end{equation}
which can be integrated to obtain the following power-law for the potential $V$ that produces solutions
\begin{equation}
    V(\chi)=V_0\chi^{n}\,,
\label{eq:V_gal}
\end{equation}
where $V_0$ is a mass scale and $n$ is the non-vanishing constant,
\begin{equation}
\label{eq:n}
    n\equiv-\frac{3\gamma}{\alpha}\,.
\end{equation}
Accordingly, Eq.~\eqref{eq:alpha} reads
\begin{equation}
\label{eq:chi_scale}
    \frac{\chi^\prime}{\chi}=-\frac{3\gamma}{n}\,.
\end{equation}

Now, we seek to determine the corresponding form of $U(\phi)$ required for the existence of scaling solutions. To do so, from Eqs.~\eqref{eq:diff} and \eqref{eq:dot_phi} we note,
\begin{equation}
    \frac{d\ln(\chi\phi'H)}{dN}=\frac{\chi'}{\chi}+\frac{\phi''}{\phi^\prime}+\frac{H'}{H}=-3\gamma\,,
\end{equation}
which leads to
\begin{equation}
\label{eq:phiprime}
    \frac{\phi''}{\phi'}=3\gamma\frac{2-n}{2n}\,,
\end{equation}
where we used Eqs.~\eqref{eq:H_scaling} and \eqref{eq:chi_scale}.

According to the latter expression, the way $\phi$ evolves depends on the value of $n$. For ${n=2}$, $\phi'$ is constant and $U(\phi)$ is an exponential potential, which corresponds to a canonical scalar field. For $n\neq2$, one finds it is $\phi^\prime/\phi$ that is constant, in which case $U(\phi)$ is a power-law form that achieves scaling. The exponential and power-law cases are presented in the next two sections, respectively.

\subsection{Canonical scalar field}
\label{sec:canonical}

By setting ${n=2}$, we can readily identify a trivial scaling solution corresponding to the canonical scalar field with an exponential potential. Following Eq.~\eqref{eq:phiprime}, $\phi''=0$ and therefore
\begin{equation}
    \phi^\prime=\lambda\,,
\end{equation}
where $\lambda$ is a non-vanishing constant. Therefore, the potential $U(\phi)$ must take an exponential form because
\be
\frac{U_{,\phi}}{U}=\frac{U'}{\phi'U}=-\frac{3\gamma}{\lambda}\,,
\ee
as ${U'/U=-3\gamma}$ according to Eq.~\eqref{eq:same_rate}. We can thus write
\be
\label{eq:U_exp}
U(\phi)=U_0\exp\left(-\frac{3\gamma}{\lambda}\phi\right)\,,
\ee
where $U_0$ is a constant energy scale. Since we are in the case $n=2$, the potential $V(\chi)$ in Eq.~\eqref{eq:V_gal} reads
\be
\label{eq:V_square}
V(\chi)=V_0\chi^2\,,
\ee
where we can set $V_0=1/2$. Then, the equation of motion for $\phi$ \eqref{eq:dot_phi} reads
\be
\label{eq:chi_square}
\dot{\phi}=\chi\,.
\ee

Plugging Eqs.~\eqref{eq:U_exp}, \eqref{eq:V_square} and \eqref{eq:chi_square}  into the hybrid three-form Lagrangian \eqref{eq:hybrid} we obtain
\be
\mathcal{L}=X-U\,,
\ee
where
\be
X = -\frac{1}{2}(\nabla\phi)^2\,,
\ee
which corresponds precisely to the Lagrangian of a quintessence field with an exponential potential, whose cosmological scaling behavior is well established~\cite{Copeland:1997et}.

\subsection{Two related power-laws}
\label{sec:generalization}

For $n\neq2$, the potential $U(\phi)$ is determined using
\be
    \frac{U'}{U}=\frac{\phi'}{\phi}\frac{\phi U_{,\phi}}{U}\,,
\ee
which leads to
\be
    \frac{\phi U_{,\phi}}{U}=\frac{2n}{n-2}\,,
\ee
in accordance with Eqs.~\eqref{eq:same_rate} and \eqref{eq:phiprime}.
Hence, integrating this expression, the potential $U$ takes a power-law form,
\be
\label{eq:U_power_law}
U(\phi)=U_0\phi^{2n/(n-2)}\,.
\ee

As a general result, plugging Eqs.~\eqref{eq:V_gal} and \eqref{eq:U_power_law} into \eqref{eq:hybrid}, we find that the three-form Lagrangian that yields scaling with the dominant background fluid consists of two power-laws whose exponents are related:
\be
\label{eq:action_scaling1}
\mathcal{L}=V_0(n-1)\chi^{n}-U_0\phi^{2n/(n-2)}\,,
\ee
with $n\neq0,1,2$.

For convenience, we can normalize the amplitude $V_0$ to unity, and introduce the constant mass scale $M$ for the potential $U$. This can be done without loss of generality, since a rescaling of the fields can always absorb the original coefficients of the two potentials. The corresponding scaling Lagrangian is
\be
\label{eq:action_scaling}
\mathcal{L}=(n-1)\chi^{n}-\left(\frac{\phi}{M}\right)^{2n/(n-2)}\,.
\ee

While it is convenient to work with this hybrid formulation, expressed by the dual variables $(\chi,\phi)$, the theory can be written in the usual three-form framework, by using Eq.~\eqref{eq:dot_chi},
\be
\mathcal{L}\propto \chi^n-\left(\dot{\chi}+3H\chi\right)^{2n/(n+2)}\,.
\ee
It can equivalently be reformulated in a pure scalar field representation, from Eq.~\eqref{eq:dot_phi},
\be
\label{eq:pure_scalar}
\mathcal{L}\propto\dot{\phi}^{n/(n-1)}-\phi^{2n/(n-2)}\,.
\ee
In Appendix \ref{app:k-Lagrangian}, we demonstrate that the latter Lagrangian is consistent with the class of scaling solutions obtained for general k-essence models~\cite{Piazza:2004df,Tsujikawa:2004dp}.

It is important to note, however, that although three-form theories can, in principle, be dualized into a scalar field, the procedure often becomes algebraically intractable \cite{Duff:1980qv,Koivisto:2009sd,SravanKumar:2016biw}. In contrast, the three-form formulation~\eqref{eq:action_scaling}, written in terms of the dual variables $(\chi,\phi)$, provides a more transparent and tractable description of the cosmological dynamics, with first-order equations of motion, Eqs.~\eqref{eq:dot_chi} and~\eqref{eq:dot_phi}, that greatly simplify the analysis.

In this context, the following section focuses on a dynamical analysis of the cosmological equations derived from the hybrid Lagrangian that provides for scaling solutions. We aim to identify the critical points in phase space and investigate the existence of cosmological attractors.

\section{Dynamical system analysis}
\label{sec:system}
\subsection{Autonomous equations}

We turn to the study of the cosmological dynamics of the model admitting scaling solutions, derived from the three-form Lagrangian in Eq.~\eqref{eq:action_scaling1}, where the potentials take the power-law form
\begin{equation}
\label{eq:power}
V(\chi)=V_0|\chi|^n\,,\quad\quad\quad U(\phi)=U_0\phi^{2n/(n-2)}\,,
\end{equation}
which permit negative $\chi$ values. The exponent $n=2$ is discarded since it corresponds to a scalar field with an exponential potential whose scaling behavior has already been extensively studied in the literature (e.g. in Ref.~\cite{Copeland:1997et}).

In addition to the three-form component, we recall that the spatially flat FLRW background also contains a cosmological fluid characterized by an adiabatic index $\gamma$, and the Friedmann equation is given by Eq.~\eqref{eq:friedmann}. Dividing it by $H^2$ naturally motivates the introduction of the dimensionless variables
\begin{equation}
\label{eq:variables}
    x^2\equiv\frac{V}{3H^2}\,,\hspace{0.5cm}y^2\equiv\frac{U}{3H^2}\,,\hspace{0.5cm}z^2\equiv\frac{\rho_\gamma}{3H^2}\,,
\end{equation}
so that the Friedman constraint becomes
\begin{equation}
    1=x^2+y^2+z^2\,,
\end{equation}
which allows us to eliminate the variable $z$. In view of Eq.~\eqref{eq:power}, the power-law indices can be expressed as
\begin{equation}
    n=\frac{\chi V_{,\chi}}{V}\,,\quad\quad\quad k\equiv\frac{2n}{n-2}=\frac{\phi U_{,\phi}}{U}\,.
\end{equation}
Furthermore, we introduce the auxiliary dimensionless variable $s$ as
\begin{equation}
\label{eq:def_s}
s\equiv\frac{\chi\phi}{H}\,,
\end{equation}
which allows us to write the system in an autonomous and tractable form. The corresponding first-order equations governing the cosmological dynamics are then given by
\begin{subequations}
\label{eq:dynamical_system1}
\begin{align}
    x'&=-\frac{3}{2}n\left(1+k\frac{y^2}{s}\right)x-\frac{H'}{H}x\,,\\
    y'&=\frac{3}{2}n\,k\frac{x^2}{s}y-\frac{H'}{H}y\,,\\
    s'&=-3\left(s+k\,y^2-n\,x^2\right)-\frac{H'}{H}s\,,
\end{align}
\end{subequations}
where
\be
\label{eq:HlH}
\frac{H'}{H}=-\frac{3}{2}\left[n\,x^2+\gamma\left(1-x^2-y^2\right)\right]\,.
\ee

The fractional energy density and equation-of-state parameter of the three-form are given by
\begin{align}
    \Omega_A&=x^2+y^2\,,\\
    w_A&=-1+\frac{n\,x^2}{x^2+y^2}\,.
\end{align}
The effective equation of state, $w_\mathrm{eff}$, describing the combined cosmological fluid (the three-form plus the background component) reads 
\begin{equation}
    w_\mathrm{eff}=-1+\gamma\left(1-x^2-y^2\right)+n\,x^2\,.
\end{equation}
Consequently, the Raychaudhuri equation is compactly written as
\begin{equation}
    \frac{H'}{H}=-\frac{3}{2}\left(1+w_\mathrm{eff}\right)\,.
\end{equation}

The dimensionality of the dynamical system can be reduced by noting that the auxiliary variable in Eq.~\eqref{eq:def_s}, ${s=\chi\phi/H}$, reads
\begin{equation}
\label{eq:s_expression}
    s=\sqrt{3}Mx^{2/n}y^{2/k}\,,
\end{equation}
where we have used $\chi$ and $\phi$ from Eq.~\eqref{eq:power}, $x^2$ and $y^2$ from Eq.~\eqref{eq:variables}, and the mass scale $M$ is defined as
\begin{equation}
\label{eq:M}
    M=V_0^{-1/n}U_0^{(2-n)/2n}\,.
\end{equation}
As a reminder, normalizing ${V_0=1}$ leads to the Lagrangian in Eq.~\eqref{eq:action_scaling} written with $\phi/M$ and the mass scale ${M=U_0^{(2-n)/2n}}$.

By substituting \eqref{eq:s_expression} into the dynamical system \eqref{eq:dynamical_system1}, the variable $s$ can be eliminated, reducing the system to two dimensions:
\begin{subequations}
\label{eq:dynamical_system2}
\begin{align}
    x'&= \frac{\sqrt{3} n }{M\left( n  -2 \right)} \left[ 2 - n \left(1 + 2 x^{-2/n} y^{(2 + n)/n} \right) \right]x \\ \nonumber
    &-\frac{H'}{H} x
\,,\\
    y'&=\frac{\sqrt{3} n^2}{M(n-2)} x^{2 - 2/n} y^{2/n}-\frac{H'}{H} y
\,,
\end{align}
\end{subequations}
where $H'/H$ is given in Eq.~\eqref{eq:HlH}.

However, the fixed points of the full three-dimensional system \eqref{eq:dynamical_system1} are more straightforward to analyze, as we shall see in the next section.

\subsection{Fixed points}
\label{sec:fixed-points}

\begin{figure*}[t]
\centering
\includegraphics[height=0.45\linewidth]{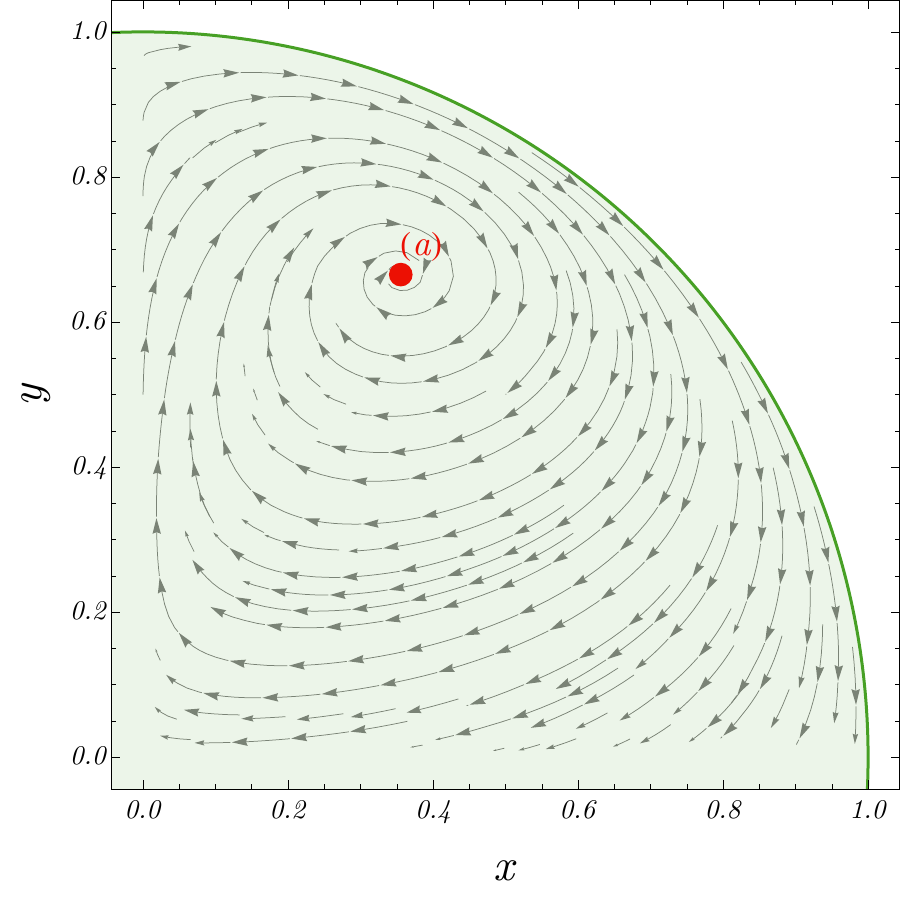}
\includegraphics[height=0.45\linewidth]{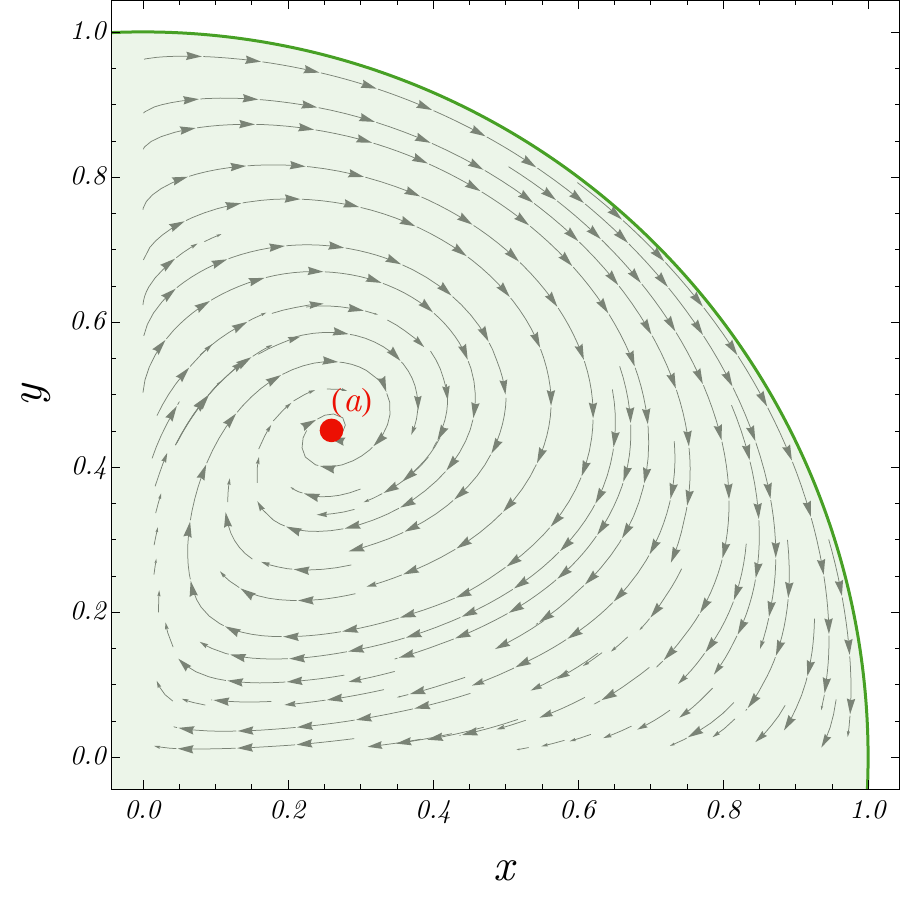}
    \caption{Phase portrait of the dynamical system showing the scaling fixed point~(a) in red, for $M^2<M_a^2$. 
Its coordinates are given by Eqs.~\eqref{eq:coord_x} and \eqref{eq:coord_y}. 
The physical region is the interior of the green arc-circle $x^{2} + y^{2} < 1$. 
Left panel: $n = 6$ and $\gamma = 4/3$. 
Right panel: $n = 4$ and $\gamma = 1$.}
    \label{fig:streamplot}
\end{figure*}

As the fixed points describe steady-state configurations, they are obtained by setting the left-hand sides of the autonomous equations~\eqref{eq:dynamical_system1} to zero and solving the resulting algebraic system. We discard the trivial solution located at the origin of the phase space, as it corresponds to the absence of the three-form field.

The dynamical system admits three nontrivial critical points: Point $(a)$, corresponding to the scaling regime, and Point $(b)$  and Point $(c)$, which describe three-form-dominated configurations. Since these points provide valuable information on the asymptotic behavior of the background cosmology, we list in Table~\ref{tab:fixed_points} the corresponding cosmological quantities associated with the three-form, namely its fractional energy density $\Omega_A$ and equation-of-state parameter $w_A$. More details are provided in Table \ref{tab:fixed_points1} of Appendix \ref{app:eigenvalues}. It is worth noting that every fixed point is characterized by
\be
\label{eq:same_s}
s=-k\,\Omega_A\,,
\ee
where the auxiliary variable $s$ is given by Eq.~\eqref{eq:def_s}, and $k=2n/(n-2)$.

\begin{table}[h!]
\renewcommand{\arraystretch}{1.4}
\centering

\begin{tabular}{c c c c}
\hline\hline
Point &  $\Omega_A$ & $w_A$ \\
\hline\\[-12pt]
$(a)$ 
& $\displaystyle \frac{3M^2}{k^2}\frac{\gamma}{n}\!\left(\frac{n}{\gamma}-1\right)^{2/k}$ 
& $-1+\gamma$ \\

$(b)$ 
& $1$ 
& $-1+nx_b^2$ \\

$(c)$
& $1$ 
& $-1+nx_c^2$ \\
\hline
\end{tabular}

\caption{
Cosmological parameters of the critical points. $x_b$ and $x_c$ are solutions of Eq.~\eqref{eq:x_b21}, and $k=2n/(n-2)$.
}
\label{tab:fixed_points}
\end{table}

We find that whenever Point~$(a)$ exists, it always corresponds to a stable scaling attractor. Point~$(b)$ is the only three-form-dominated configuration that can be stable, and this occurs exclusively when the scaling solution is absent. Point~$(c)$ is always unstable. The detailed conditions governing the existence and stability of the three fixed points are presented in Appendix~\ref{app:eigenvalues}. They are listed in Table~\ref{tab:existence_stability}. Below we summarize the key features that are most relevant for the cosmological dynamics.

\textbf{Point~$(a)$:} This is the critical point that corresponds to the scaling of the three-form component with the background fluid, where the three-form energy density parameter and equation of state take the values
\begin{align}
\label{eq:Omega_scaling1}
    \Omega_A&=\frac{3M^2}{k^2}\frac{\gamma}{n}\left(\frac{n}{\gamma}-1\right)^{2/k}\,,\\
    w_A&=-1+\gamma\,,
\end{align}
respectively, where ${k=2n/(n-2)}$ and the mass scale $M$ was defined in Eq.~\eqref{eq:M}. At the scaling we find that the two potentials scale proportionally, as expected,
\be
\label{eq:UoverV1}
\frac{U}{V}=\frac{n}{\gamma}-1\,.
\ee

The scaling point exists provided that $0<\Omega_A<1$, that is when the following two conditions hold together,
\be
\label{eq:scaling_exists}
n>\gamma\quad\mathrm{and}\quad M^2 < M_a^2\,,
\ee
where
\be
\label{eq:S_a1}
M_a^2=\frac{k^2}{3}\frac{n}{\gamma}\left(\frac{n}{\gamma}-1\right)^{-2/k}\,,
\ee
from Eq.~\eqref{eq:Omega_scaling1}, with $k=2n/(n-2)$.

As for its stability, the analysis of the eigenvalues derived from the Jacobian matrix at Point~$(a)$, listed in Table~\ref{tab:eigenvalues} of Appendix \ref{app:eigenvalues}, shows that, provided it exists, the scaling point is always an attractor, either as a stable spiral or as a stable node. Figure~\ref{fig:streamplot} gives one illustration of the phase space trajectories corresponding to the radiation and matter scaling regimes for ${(n,\gamma)=(6,4/3)}$ and ${(n,\gamma)=(4,1)}$, respectively.

\textbf{Point~$(b)$:} This equilibrium configuration corresponds to a universe fully dominated by the three-form field, for which the density and equation-of-state parameters are given by
\begin{align}
\label{eq:Omega_domination1}
    \Omega_A&=1\,,\\
    w_A&=-1+nx_b^2\,,
\label{eq:w_domination1}
\end{align}
where
\be
w_A\leqslant1\,, \quad(x_b^2\leqslant2/n\quad\mathrm{or}\quad n<0)\,,
\ee
and $x_b^2$ satisfies the following algebraic equation, derived in Appendix \ref{app:eigenvalues}, from Eqs.~\eqref{eq:s_expression} and \eqref{eq:same_s},
\be
\label{eq:x_b21}
(x^2)^{2/n}(1-x^2)^{2/k}=\frac{k^2}{3M^2}\,,
\ee
where $k=2n/(n-2)$. From Eq.~\eqref{eq:w_domination1}, we notice that the three-form fluid behaves as a phantom dark energy component for ${n<0}$.

In terms of stability, the eigenvalues at Point~$(b)$ are both negative only if the conditions in Eq.~\eqref{eq:scaling_exists} do not hold, 
\be
n<\gamma\quad\mathrm{or}\quad M^2 > M_a^2\,,
\ee
where $M_a^2$ is given in Eq.~\eqref{eq:S_a1}. Point~$(b)$ is thus an attractor if the scaling point is absent. Otherwise, this three-form dominated configuration does not exist or is a saddle point.

\begin{figure*}[t]
\centering
\includegraphics[height=0.45\linewidth]{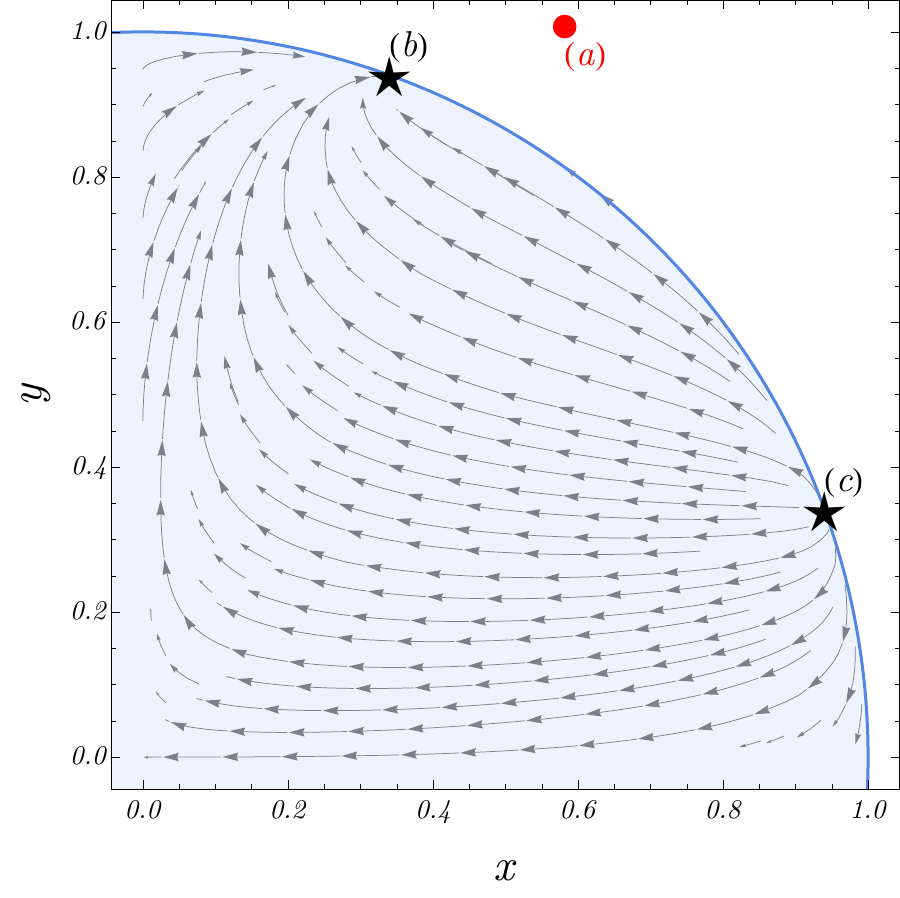}
    \caption{Phase portrait for $\gamma=1$ and $n=4$. For $M^2>M_a^2$, the scaling Point~$(a)$ becomes nonphysical, and the dynamical system is attracted toward the three-form dominated configuration in Point $(b)$. The other three-form domination Point $(c)$ is a repeller.}
    \label{fig:boundary1}
\end{figure*}

\textbf{Point~$(c)$:} As the latter, this fixed-point also corresponds to  a universe fully dominated by the three-form field satisfying
\begin{align}
\label{eq:Omega_domination2}
    \Omega_A&=1\,,\\
    w_A&=-1+nx_c^2\,,
\label{eq:w_domination2}
\end{align}
where
\be
w_A>1\,, \quad(x_c^2>2/n)\,,
\ee
and $x_c^2$ is the second solution of Eq.~\eqref{eq:x_b21}. Point~$(c)$ is unable to accelerate the universe and is an instable node that repels all trajectories, as illustrated in Fig.~\ref{fig:boundary1}.

\begin{figure*}[t]
\centering
\includegraphics[height=0.4\linewidth]{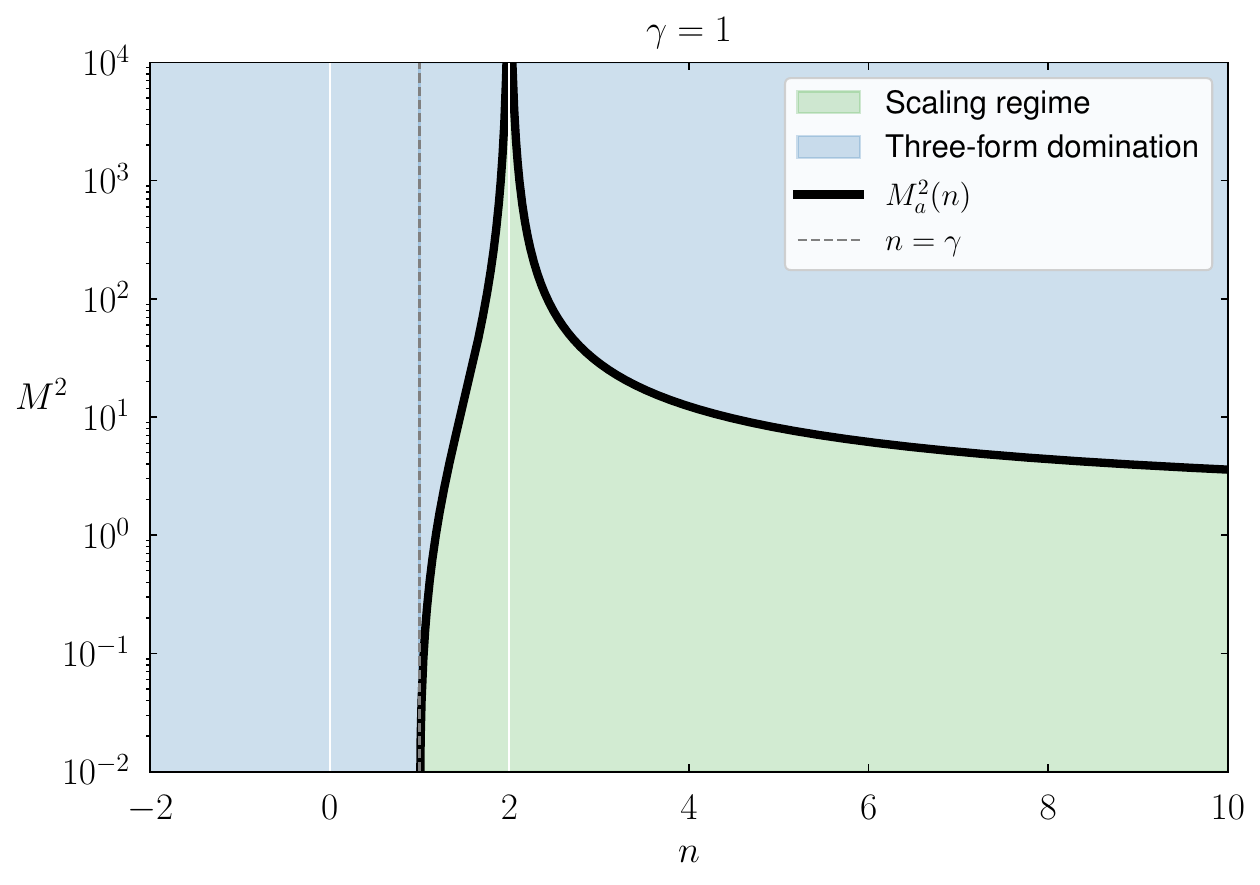}
    \caption{Region of stability of the scaling regime and three-form-dominated universe, in the parameter space $(n,S^2)$, for $\gamma=1$. The values $n=(0,2)$ are excluded. The boundary $M_a^2$ between the two attractors is given by Eq.~\eqref{eq:S_a1}.}
    \label{fig:regions}
\end{figure*}

For clarity, Fig.~\ref{fig:regions} illustrates the conditions governing the two possible stable configurations: either Point~$(a)$ for the scaling or Point~$(b)$ for the three-form domination, in the parameter space $(n,M^2)$, for a dust background fluid ($\gamma=1$).

It is worth noting that when the dynamical system lies in the scaling regime, its stability prevents any transition toward three-form domination. In other words, once the system enters the scaling attractor, it remains in that state. One possibility for achieving a viable cosmological evolution, where a dark-energy three-form dominates at late times, capable of exiting the initial scaling with the background fluid, is to introduce additional terms in the Lagrangian~\eqref{eq:action_scaling}. We will explore this path in the following section.

\section{Scaling exit toward cosmic acceleration}
\label{sec:exit}

\subsection{Cosmological constant solution}

The simplest realization of a dark energy component is to add a cosmological constant to the hybrid Lagrangian~\eqref{eq:hybrid}.
However, this approach does not exploit the intrinsic dynamics of the three-form field to generate dark energy following the scaling behavior.
A similar limitation arises if we include an exponential potential term, function of $\phi$, to induce a late-time transition from the scaling solution toward cosmic acceleration. The effect of this extra term remains essentially equivalent to that of a cosmological constant.

Another possibility is to consider a Lagrangian density of the form
\begin{equation}
    \mathcal{L}\propto-\left(\phi^k+\phi^{-k}\right)+\chi^n\,,
\end{equation}
where $k=2n/(n-2)$. This model triggers the exit from the scaling regime at late times because it is a saddle point in the associated dynamical system, which can be straightforwardly adapted from Sec.~\ref{sec:system}. The trajectories are eventually attracted toward a stable three-form–dominated node. Nevertheless, we observe that this exit solution still remains close to a cosmological-constant–like behavior, as the field $\phi$ asymptotically vanishes while undergoing damped oscillations.

\subsection{Double potential structure}

Alternatively, we want to introduce a scenario that enables a transition out of the scaling regime toward the domination of dark energy, which could be distinguished from a cosmological constant. We propose to include additional terms to the scaling three-form Lagrangian while preserving its original structure, thereby avoiding unnecessary parameter proliferation and degeneracies. Specifically, we extend both potentials as
\begin{align}
\label{eq:double_V}
    V(\chi)&=V_0\chi^n+V_1\chi^m\,,\\
\label{eq:double_U}
    U(\phi)&= U_0\phi^{2n/(n-2)}+U_1\phi^{2m/(m-2)}\,,
\end{align}
introducing the additional parameter $m$ and mass scales $V_1$ and $U_1$. When we build the corresponding working model in Section \ref{sec:toy_model}, these additional mass scales help us reproduce today's cosmology with matter abundance $\Omega_m^0\simeq0.3$.

The idea is to construct two dynamical systems of the same form as that analyzed in Sec.~\ref{sec:system}, exploiting the scaling point of the first and the three-form-dominated attractor of the second. This is inspired by quintessence models with exponential potentials, in which a double-exponential structure enables both scaling behavior and cosmic acceleration. Indeed, by combining potentials of different slopes, $\alpha$ and $\beta$,
\begin{equation}
    V(Q)\propto e^{\alpha Q}+e^{\beta Q}\,,
\end{equation}
for a scalar field $Q$, it is possible to take advantage of both attractors within a unified dynamical framework~\cite{Barreiro:1999zs}. Appropriate values of the first slope realize a period of scaling, while the second evolves the solution towards scalar field domination. Remarkably, the one-parameter scalar-field model first proposed in Ref.~\cite{Nunes:2003ff}
\begin{equation}
\label{eq:double}
    V(Q)\propto e^{-\frac{3}{\lambda} Q}+e^{-\lambda Q}\,,
\end{equation}
is fully characterized by a single parameter, $\lambda$, and admits an analytic solution for the dark-energy density (in Eq.~\eqref{eq:solution_rho_A} in the next section) which makes the model particularly convenient for cosmological studies \cite{daFonseca:2021imp,daFonseca:2022qdf,Barros:2022kpo,daFonseca:2023ury,CosmoVerseNetwork:2025alb}. It accounts for both early-time scaling and late-time acceleration, yielding a form of dark energy component that is dynamically distinct from a cosmological constant.

In the following section, we construct a viable family of scaling-exit solutions within our three-form framework, which is analog to this one-parameter model. Although the scalar double-exponential picture motivates our approach, the formalism is different and rely on the double-potential structure introduced in Eqs.~\eqref{eq:double_V} and \eqref{eq:double_U}. The leading terms $\chi^n$ and $\phi^{2n/(n-2)}$ reproduce the scaling behavior with matter, corresponding to the critical Point $(a)$ of the associated dynamical system. At late times, the initially subdominant contributions $\chi^m$ and $\phi^{2m/(m-2)}$ become dominant, driving the system toward the three-form–dominated configuration at the critical Point~$(b)$. In this framework, the transition toward dark-energy domination emerges dynamically, without introducing arbitrary potential structures. 

\subsection{Working model}
\label{sec:toy_model}

\begin{figure*}[t]
\centering
\includegraphics[width=0.49\textwidth]{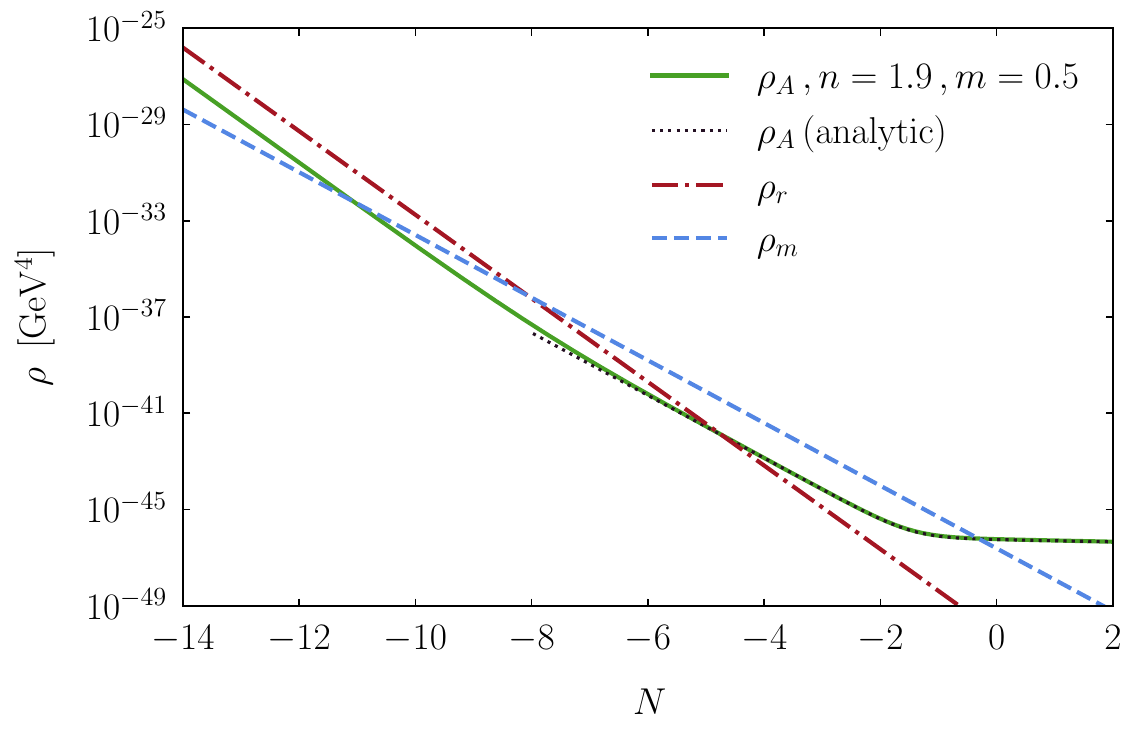}
\includegraphics[width=0.49\textwidth]{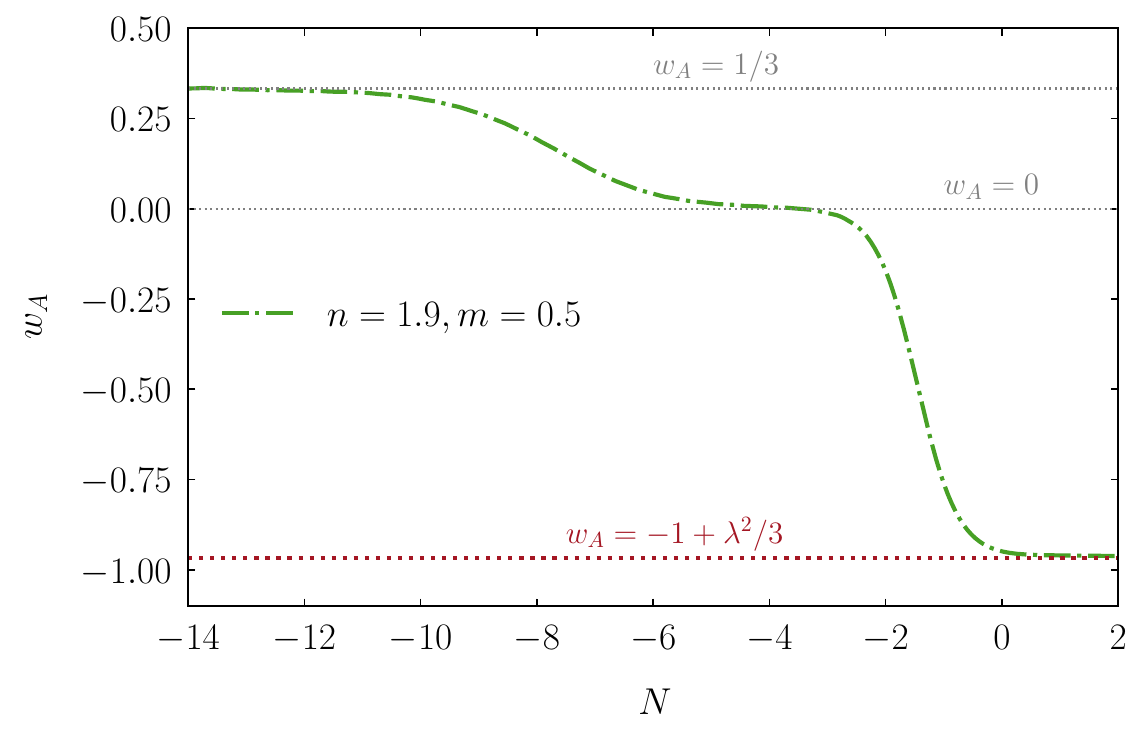}
    \caption{Left panel: Evolution of the energy densities of radiation $(\rho_r)$, matter ($\rho_m$), and the three-form $(\rho_A)$, for $\lambda^2=0.1$, $n=1.9$, and $m=0.5$. The black dotted curve corresponds to the analytic solution of $\rho_A$ in Eq.~\eqref{eq:solution_rho_A} during the matter-dominated era. Right panel: Evolution of the three-form equation of state $(w_A)$ for the same parameters.}
    \label{fig:scaling}
\end{figure*}

To build a working cosmological model, let us begin with the continuity equation for the three-form field,
\begin{equation}
    \label{eq:continuity}
    \rho_A^\prime+3\left(\rho_A+p_A\right)=0\,,
\end{equation}
where $\rho_A+p_A=\chi V_{,\chi}$, according to Eqs.~\eqref{eq:rho_A} and \eqref{eq:p_A}.
This allows the continuity equation to be rewritten as
\be
\label{eq:continuity2}
\rho_A^\prime+3H^2\frac{\chi V_{,\chi}}{H^2}=0\,,
\ee
using the Friedmann constraint $3H^2=\rho_A+\rho_m$, where $\rho_m$ denotes the matter density, assuming it dominates over radiation $(\gamma=1)$. We introduce the parameter $\lambda$ defined as
\be
\label{eq:lambda}
\lambda^2\equiv\frac{\chi V_{,\chi}}{H^2}=s\frac{\phi^\prime}{\phi}\,,
\ee
recalling Eq.~\eqref{eq:def_s}. The continuity equation \eqref{eq:continuity2} then becomes
\begin{equation}
    \label{eq:continuity3}
    \rho_A^\prime+\left(\rho_m+\rho_A\right)\lambda^2=0\,.
\end{equation}

The analytical solution of this differential equation is known from the case of quintessence with a double-exponential potential in Eq.~\eqref{eq:double}, under the assumption that $\lambda^2$ remains constant throughout the evolution~\cite{Nunes:2003ff},
\begin{equation}
    \label{eq:solution_rho_A}
    \rho_A=\rho_m^0\left[\left(\frac{1-\Omega_m^0}{\Omega_m^0}-\frac{\lambda^2}{3-\lambda^2}\right)e^{-\lambda^2N}+\frac{\lambda^2}{3-\lambda^2}e^{-3N}\right]\,,
\end{equation}
where $\rho_m^0=3H_0^2\Omega_m^0$ is today's matter density, and $H_0$ is the Hubble constant. This solution is especially noteworthy because the single parameter $\lambda^2<2$ simultaneously determines both the level of scaling in the matter attractor,
\be
\Omega_A=\frac{\lambda^2}{3}\,,\quad w_A=0\,,
\ee
and the late-time dark-energy equation of state in the accelerated attractor,
\be
\label{eq:late_time}
\Omega_A=1\,, \quad w_A=-1+\frac{\lambda^2}{3}\,.
\ee
Importantly, the asymptotic value of $w_A=-1+\lambda^2/3$ differs from that of a cosmological constant, while the $\Lambda$CDM limit is recovered for $\lambda \rightarrow 0$.

In our three-form construction, it is the leading terms,
\begin{equation}
    \chi^n\quad\mathrm{and}\quad\phi^{2n/(n-2)}\,,
\end{equation}
that generate the scaling regime at Point~$(a)$. Using Eqs.~\eqref{eq:phiprime}, \eqref{eq:same_s} and \eqref{eq:Omega_scaling1} in Eq.~\eqref{eq:lambda}, we find that the attractor value of $\lambda^2$ is  
\be
    \label{eq:lambda3} 
    \lambda^2=\frac{9M^2}{4n^3}(n-2)^2(n-1)^{(n-2)/n}\,,
\ee
where $M$ is defined in Eq.~\eqref{eq:M}.
This expression confirms a degeneracy between the exponent $n$ and the mass scale $M$: different pairs $(n,M)$ yield the same $\lambda^2$ and hence the same background scaling evolution.

At late times, we require that the previous subdominant terms,
\begin{equation}
    \chi^m\quad\mathrm{and}\quad\phi^{2m/(m-2)}\,,
\end{equation}
govern the evolution and lead the system to the three-form–dominated Point~$(b)$. Using Eq.~\eqref{eq:w_domination1}, with the exponent $m$, combined with Eq.~\eqref{eq:late_time}, we find the coordinate of Point~$(b)$,
\begin{equation}
\label{eq:xb_m}
x_b^2 = \frac{\lambda^2}{3m}\,.
\end{equation}
Similarly to Eq.~\eqref{eq:M}, introducing
\begin{equation}
    M_1=V_1^{-1/m}U_1^{(2-m)2/m}\,,
\end{equation}
and solving Eq.~\eqref{eq:x_b21} for the mass scale $M_1$, we obtain
\begin{equation}
    \label{eq:sol_m}
    M_1^2=\frac{4m^3}{\lambda^2(m-2)^2}\left(\frac{3m}{\lambda^2}-1\right)^{(2-m)/m}\,,
\end{equation}
which ensures that the same $\lambda^2$ governs both eras, for a given exponent $m$. Different combinations of $(U_1,V_1)$ lead to the same asymptotic three-form–dominated attractor $({\Omega_A\rightarrow1})$ since the right-hand side of Eq.~\eqref{eq:sol_m} is invariant under the rescaling
\be
\label{eq:rescale_Ua}
U_1 \to \alpha\,U_1\,, \quad V_1 \to \alpha^{(2-m)/2} V_1\,,
\ee
for any $\alpha>0$. However, different values of $\alpha$ lead to different cosmological evolutions along the trajectory toward this attractor. In particular, reproducing the present-day matter abundance, because it lies in-between attractors, requires fine tuning $(U_1,V_1)$.

To verify the full dynamical behavior of our working model, we numerically integrate the equations of motion for $\chi$ and $\phi$, Eqs.~\eqref{eq:dot_chi} and \eqref{eq:dot_phi}, together with the background, including radiation. Initial conditions are chosen in the radiation-era attractor. For illustrative purpose, Fig.~\ref{fig:scaling} shows an example for
\be
{\{\lambda^2=0.1,n=1.9,m=0.5\}}\,.
\ee
In the left panel, $\rho_A$ tracks radiation and then matter during their respective domination epochs. At late times, it overtakes matter and drives the accelerated expansion. Correspondingly, the right panel shows $w_A$ evolving from $1/3$ to $0$ and finally to $-1+\lambda^2/3$.

\begin{figure}[h!]
\includegraphics[height=0.75\linewidth]{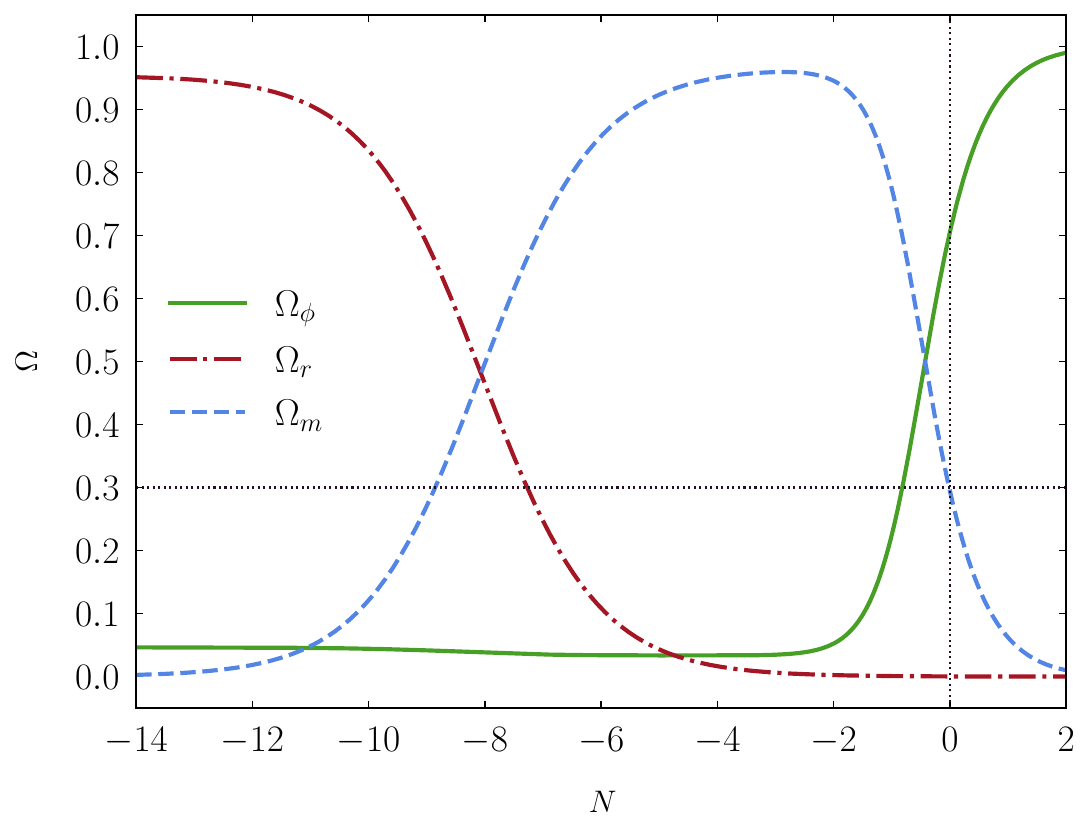}
    \caption{Evolution of the fractional energy densities of radiation $(\Omega_r)$, matter $(\Omega_m)$ and the three-form $(\Omega_A)$, for $\lambda^2=0.1$, $n=1.9$, and $m=0.5$.}
    \label{fig:abundances}
\end{figure}

The left panel of Fig.~\ref{fig:scaling} demonstrates that the numerical results (green solid line) agree with the analytic prediction (black dotted line) throughout the matter-dominated scaling regime and the three-form–dominated epoch. The present-day value of $\Omega_m^0 = 0.3$ is successfully recovered, as shown in Fig.~\ref{fig:abundances}.

Several conditions must be satisfied in order to obtain a viable cosmology. 
First, the early-time scaling regime must exist, which constrains the allowed range of the exponent $n$ and the associated mass scale $M$. 
For a given choice of $n$, one can also infer observational bounds on $M$ by exploiting existing constraints on the parameter $\lambda$ obtained in previous studies~\cite{daFonseca:2021imp}. 
Solving Eq.~\eqref{eq:lambda3} for $M$, a $1\sigma$ upper bound $\lambda<\lambda_{1\sigma}$ immediately implies
\begin{equation}
    M^2<\lambda_{1\sigma}^2\,
    \frac{4n^3}{9\,(n-2)^2\,(n-1)^{(n-2)/n}}\,.
\end{equation}
For our numerical example, $n=1.9$, the early universe $1\sigma$ constraint $\lambda<0.056$ yields
\begin{equation}
    M^2<0.95\,.
\end{equation}
Local universe constraints are significantly weaker. Using $\lambda=0.50\pm0.24$ at $1\sigma$, and adopting the conservative upper bound $\lambda<0.74$, we obtain instead
\begin{equation}
    M^2<1.66\times 10^2\,.
\end{equation}

Second, the late-time exponent must obey $m < n$, ensuring that the second dynamical system eventually dominates and drives the expansion. A simple sufficient condition for the existence of the corresponding attractor is $m < 1$: in this case the late-time system does not admit a scaling regime, and the three-form–dominated point is automatically stable (see Fig.~\ref{fig:regions}).

We also have to consider the sign and evolution of $\chi$. For $n > 2$ and $m < 1$, it typically crosses zero, causing $V_{,\chi} \propto |\chi|^{m - 1}$ to diverge at late times and resulting in model instability. Conversely, for $n > 2$ and $m > 1$, $\chi$ tends to oscillate around zero, yielding $w_A \to -1$. The resulting cosmology becomes observationally indistinguishable from a cosmological constant, defeating the purpose of the proposed framework. A good choice is therefore to choose exponents that satisfy the condition $n<2$ (and $m<1$) like in our illustration.

Finally, the model requires some degree of fine-tuning of the mass scales, which are degenerate with the power-law exponents governing the scaling regime and the onset of late-time acceleration. This is necessary to reproduce the observed cosmology, and it is akin to that encountered in quintessence models.

\section{Conclusion}\label{sec:conclusions}

Starting from a first-order action for a massive three-form field, we have constructed an original {\it hybrid} representation in which its dynamics can be recast in terms of two dual scalar degrees of freedom. This formulation provides a tractable framework to identify scaling solutions in three-form dark energy, and has allowed us to derive general conditions that the underlying Lagrangian must satisfy in order for the three-form energy density to scale with the dominant background fluid. Our dynamical-system analysis further shows that these scaling regimes are attractors, leading to robust cosmological solutions that are weakly sensitive to initial conditions. The analysis also shows the existence of three-form-dominated universe configurations that are stable.

In the minimal setup, however, the scaling solution remains stable, preventing the system from naturally evolving toward a late-time three-form–dominated accelerated phase. Achieving a transition from early-time scaling to late-time acceleration therefore requires extending the Lagrangian. Simple extensions that effectively mimic a cosmological constant are of limited interest, as they freeze the field dynamics and reduce the scenario to a nearly non-dynamical dark-energy component. Instead, we have identified a more compelling construction based on a double potential structure that preserves the original functional form of the hybrid Lagrangian.

Overall, under appropriate conditions and fine-tuning, this mechanism provides a viable realization of scaling dark energy in three-form cosmology. In particular, it gives rise to a late-time acceleration phase that remains distinct from a pure cosmological constant, while remaining compatible with standard dark energy phenomenology.


\acknowledgments
We thank Cristiana Pinto, Rafael Pastor, Sara Dur\~{a}o, Lucas Habib for discussions during the first phases of this work.
It is supported by Funda\c{c}\~{a}o para a Ci\^{e}ncia e a Tecnologia (FCT) through the research grants DOI: 10.54499/UIDB/04434/2020 and DOI: 10.54499/UIDP/04434/2020. VdF acknowledges support from FCT through grant 2022.14431.BD.

\appendix

\section{Existence and stability of the critical points}
\label{app:eigenvalues}

For completeness, this appendix provides additional details on the conditions for the existence and stability of the three nontrivial fixed points~$(a)$, $(b)$, and $(c)$ of the dynamical system, discussed in Section~\ref{sec:fixed-points}. These points are listed in Table~\ref{tab:fixed_points1} for ease of reference.

\begin{table}[h!]
\centering
\renewcommand{\arraystretch}{1.3}
\setlength{\tabcolsep}{4pt}

\begin{tabular}{c|c|c|c|c|c}
\hline\hline
Point & $x^2$ & $y^2$ & $s$ & $\Omega_A$ & $w_A$ \\
\hline
$(a)$ 
& $x_a^2$ 
& $\left(\frac{n}{\gamma}-1\right)x_a^2$
& $-k\frac{n}{\gamma}x_a^2$
& $\frac{n}{\gamma}x_a^2$
& $-1+\gamma$
\\

$(b)$
& $x_b^2$
& $1-x_b^2$
& $-k$
& $1$
& $-1+n x_b^2$
\\

$(c)$
& $x_c^2$
& $1-x_c^2$
& $-k$
& $1$
& $-1+n x_c^2$
\\
\hline
\end{tabular}

\caption{
Critical points of the dynamical system.  
$k=2n/(n-2)$ and $M$ is the potential mass scale.  
$x_a^2$ is defined in Eq.~\eqref{eq:coord_x};  
$x_b^2$ and $x_c^2$ in Eq.~\eqref{eq:x_b}.  
}
\label{tab:fixed_points1}
\end{table}

To determine the nature of the critical points, we analyze the behavior of trajectories in their vicinity by introducing small perturbations around each equilibrium configuration and assessing the resulting linearized dynamics to assess the system’s stability. The corresponding eigenvalues of the Jacobian matrix, evaluated at the critical points~$(a)$, $(b)$ and $(c)$,  are summarized in Table~\ref{tab:eigenvalues} .

\begin{table*}[t]
\centering
\renewcommand{\arraystretch}{1.45}
\setlength{\tabcolsep}{10pt}

\begin{tabular}{l l}
\hline\hline
Point & Eigenvalues $(x,y)$\\
\hline
\\[-10pt]
$(a)$ &
$\displaystyle 
\lambda_{1,2}=\frac{3}{4}(\gamma-2)\pm 
\frac{3}{4}(n-2)n^2
\left[4+8n x_a^2 (n-\gamma)-\gamma(8n-9\gamma+4)\right]^{1/2}$
\\[10pt]  

$(b)$ &
$\displaystyle 
\lambda_{1,2}=\left(
-3+\frac{3n x_b^2}{2}\,,\;
3n x_b^2-3\gamma
\right)$
\\[10pt]  

$(c)$ &
$\displaystyle 
\lambda_{1,2}=\left(
-3+\frac{3n x_c^2}{2}\,,\;
3n x_c^2-3\gamma
\right)$
\\[6pt]  

\hline
\end{tabular}

\caption{
Eigenvalues of the Jacobian matrix for the critical points. 
$x_a^2$ is given by Eq.~\eqref{eq:coord_x}, and $x_b^2$, $x_c^2$ by Eq.~\eqref{eq:x_b}.
}
\label{tab:eigenvalues}
\end{table*}

\textbf{Point~$(a)$ (scaling solution).} This critical point corresponds to the scaling regime. Its coordinates are
\begin{align}
\label{eq:coord_y}
    y_a^2=&\left(\frac{n}{\gamma}-1\right)x_a^2\,\,,\\
    s_a =&-k\frac{n}{\gamma}x_a^2\,, 
    \label{eq:coord_s}
\end{align}
with $k=2n/(n-2)$. The quantity $x_a^2$ follows from Eq.~\eqref{eq:s_expression},
\be
\label{eq:coord_x}
x_a^2=\frac{3M^2}{k^2}\gamma^2\left(\frac{n}{\gamma}-1\right)^{2/k}\,,
\ee
where $M$ is the mass scale of the potentials defined in Eq.~\eqref{eq:M}. Using Eqs.~\eqref{eq:coord_y}, \eqref{eq:coord_x} and \eqref{eq:variables}, the ratio of the two potentials read,
\be
\label{eq:UoverV}
\frac{U}{V}=\frac{y_a^2}{x_a^2}=\frac{n}{\gamma}-1\,.
\ee

The fractional energy density and equation of state of the three-form field are
\begin{align}
\label{eq:Omega_scaling}
    \Omega_A&=\frac{n}{\gamma}x_a^2\,,\\
    w_A&=-1+\gamma\,,
\end{align}
respectively. Inserting Eq.~\eqref{eq:coord_x}, the energy density fraction becomes
\begin{equation}
\label{eq:Omega_A}
    \Omega_A=\frac{3M^2}{k^2}\frac{\gamma}{n}\left(\frac{n}{\gamma}-1\right)^{2/k}\,,
\end{equation}
where $k=2n/(n-2)$.

The condition $n>\gamma$ ensures that $y^2$ from Eq.~\eqref{eq:coord_y} is real. Moreover, requiring $\Omega_A<1$ implies
\be
x_a^2<\gamma/n\,.
\ee
Thus, following Eq.~\eqref{eq:coord_x}, the scaling critical point exists only when
\be
n>\gamma\quad\mathrm{and}\quad M^2 < M_a^2\,,
\ee
where
\be
\label{eq:S_a}
M_a^2=\frac{k^2}{3}\frac{n}{\gamma}\left(\frac{n}{\gamma}-1\right)^{-2/k}\,,
\ee
with $k=2n/(n-2)$.

To determine stability, we examine the eigenvalues $\lambda_{1,2}$ listed in Table~\ref{tab:eigenvalues}. The scaling point is always a stable attractor. It behaves as a stable spiral when $M^2<M_\star^2$ and a stable node when $M^2>M_\star^2$, where
\begin{equation}
\label{eq:S_*}
M^2_*=\frac{k^2}{3\gamma^2}\frac{\gamma(4 + 8n - 9\gamma) -4}{8n(n-\gamma)} \left(\frac{n - \gamma}{\gamma}\right)^{-2/k}\,.
\end{equation}
Indeed, for $M^2<M_*^2$, the discriminant of the eigenvalues is negative and both eigenvalues are complex with negative real part
\be
\frac{3}{4}(\gamma-2)<0\,,
\ee
since $\gamma<2$. Conversely, for $M^2>M_*^2$ the eigenvalues are real and negative provided $n>\gamma$ and $x_a^2<\gamma/n$, which are the conditions of existence of this critical point. Thus, Point~$(a)$ is a scaling attractor.

\begin{table*}[t]
\centering
\renewcommand{\arraystretch}{2}
\begin{tabularx}{\textwidth}{c|Y|Y!{\vrule width 1.2pt}Y|Y|Y}
\hline\hline
 & \multicolumn{2}{c!{\vrule width 1.2pt}}{$n<2$}
 & \multicolumn{3}{c}{$n>2$} \\ \cline{2-6}
Point
 & $M^{2}<M_a^{2}$
 & $M^{2}>M_a^{2}$
 & $M^{2}<M_b^{2}$
 & $M_b^{2}<M^{2}<M_a^{2}$
 & $M^{2}>M_a^{2}$ \\ \hline
$(a)$ & \textbf{stable} & -- & \textbf{stable} & \textbf{stable} & -- \\
$(b)$ & saddle & \textbf{stable} & -- & saddle & \textbf{stable} \\
$(c)$ & -- & -- & -- & unstable & unstable \\
\hline
\end{tabularx}
\caption{Existence and stability of the critical points. The mass scale thresholds $M_a$ and $M_b$ are given by Eq.~\eqref{eq:S_a} and Eq.~\eqref{eq:M_b}, respectively.}
\label{tab:existence_stability}
\end{table*}

\textbf{Points $(b)$ and $(c)$ (three-form–dominated)}: Both fixed-points correspond to a universe fully dominated by the three-form,  
\begin{align}
\label{eq:Omega_domination}
    \Omega_A&=1\,,\\
    w_A&=-1+nx_{b,c}^2\,.
\label{eq:w_domination}
\end{align}
Since at a three-form–dominated point
\ba
y^2=1-x^2\,,\quad\quad s^2 = k^2\,,
\ea
where $s$ is given by Eq.~\eqref{eq:s_expression}, $x_b^2\leqslant2/n$ and $x_c^2>2/n$ are the two solutions of
\be
\label{eq:x_b}
(x^2)^{2/n}(1-x^2)^{2/k}=\frac{k^2}{3M^2}\,,
\ee
with $k=2n/(n-2)$.

These two solutions differ cosmologically through their equation of state:
\ba
\mathrm{Point}~(b): w_A\leqslant1\,,\quad\quad \mathrm{Point}~(c): w_A>1\,.
\ea

For $n<2$, only Point~$(b)$ exists.
For $n>2$, both points exist only if the maximum of Eq.~\eqref{eq:x_b} at $x^2=2/n$ is positive, leading to the condition
\be
M^2>M_b^2\,,
\ee
with
\be
\label{eq:M_b}
M_b^2=\frac{k^2}{3}\left(\frac{2}{n}\right)^{-2/n}\left(1-\frac{2}{n}\right)^{-2/k}\,.
\ee
When $M^2=M_b^2$, Points~$(b)$ and $(c)$ coincide.

Regarding stability, for Point~$(b)$ the first Jacobian eigenvalue is always negative since $x_b^2<2/n$. The second is negative when $x_b^2<\gamma/n$. Thus, if $n<\gamma$, Point~$(b)$ is always stable because $x_b^2<1$. If $n>\gamma$, by using $x_b<\gamma/n$ we find that stability requires 
\be
M^2>M_a^2\,,
\ee
where $M_a$ is given by Eq.~\eqref{eq:S_a}, which is the same threshold that prevents the existence of the scaling point. If $M^2<M_a^2$, then $\lambda_2>0$ and Point~$(b)$ is a saddle point.

For Point~$(c)$, both eigenvalues are positive because $x_c^2>2/n$ and $\gamma<2$. Point~$(c)$ is therefore always unstable.

A summary of existence and stability conditions for each non-trivial critical point is given in Table~\ref{tab:existence_stability}.

\section{Comparison with scaling k-Lagrangians}
\label{app:k-Lagrangian}
\subsection{Finding the equivalent k-essence model}

Within the context of general k-essence models, the class of Lagrangians admitting scaling solutions is well established~\cite{Piazza:2004df,Tsujikawa:2004dp}. In this Appendix, we demonstrate the consistency between these solutions and the three-form Lagrangian \eqref{eq:action_scaling1} that exhibits scaling behavior, derived in Sec.~\ref{sec:generalization} from a first order formulation. This correspondence arises because three-form theories can be dualized into an equivalent scalar-field framework, thereby establishing the connection between the two descriptions of scaling cosmologies.

Our starting point is the three-form hybrid action~\eqref{eq:hybrid},
\be
\mathcal{L}= -U(\phi)-V(\chi)+\chi V_{,\chi}\,,
\ee
which admits a trivial scaling solution for $V=V_0\chi^2$ ($V_0=1/2$), corresponding to the canonical scalar field with an exponential potential, as derived in Sec.~\ref{sec:canonical}.

We assume the general scaling potential $V(\chi)$ in Sec.~\ref{sec:generalization},
\ba
V&=&V_0\chi^n\,,\\ \label{eq:V}
\mathcal{L}&=&-U(\phi)+V_0(n-1)\chi^n\,,\label{eq:gen}
\ea
for $n\neq0$ and $n\neq2$. We aim to determine the corresponding and most general form of the function $U(\phi)$ that yields scaling, by using the k-essence models, and to confirm that it coincides with the power law in Eq.~\eqref{eq:U_power_law} that we obtained within the three-form framework.

Following Eq.~\eqref{eqmov2}, let $X$ denote the kinetic term,
\ba
\label{eq:kinetic_term}
    X&=&-\frac{1}{2}\left(\nabla\phi\right)^2\,,\\
    &=&\frac{n^2V_0^2}{2}\chi^{2(n-1)}\,,
\ea
which can be inverted (for $n\neq1$) to yield
\begin{equation}
\label{eq:chi_X}
    \chi=\left(\frac{2X}{n^2V_0^2}\right)^{1/2(n-1)}\,.
\end{equation}
Substituting this relation into the Lagrangian \eqref{eq:gen}, we obtain
\begin{equation}
\label{eq:L0}
    \mathcal{L}=-U(\phi)+\frac{n-1}{V_0^{1/(n-1)}}\left(\frac{2X}{n^2}\right)^{n/2(n-1)}\,.
\end{equation}
    
We now map our three-form theory onto a non-canonical scalar field $\varphi$ defined as
\begin{equation}
\label{eq:varphi}
    \varphi=-\ln\phi\,,
\end{equation}
whose kinetic term, ${\tilde{X}=-(1/2)g^{\mu\nu}\partial_\mu\varphi\partial_\nu\varphi}$, reads
\begin{equation}
\label{eq:X_tilde}
    \tilde{X}=Xe^{2\varphi}\,.
\end{equation}
Defining
\begin{equation}
\label{eq:Y}
    \tilde{Y}=\tilde{X}e^{k\varphi}\,,
\end{equation}
for an arbitrary parameter $k$, we can re-write Eq.~\eqref{eq:L0} in the form of a k-essence Lagrangian,
\begin{align}
    \label{eq:P4}
    P(\varphi,\tilde{X}) &= \tilde{X} \bigg[-\frac{U}{\tilde{Y}}e^{k\varphi}+c\,\tilde{Y}^{(2-n)/2(n-1)} \\ \nonumber
    &\quad\times \exp\left({\frac{\varphi}{2}\frac{k(n-2)-2n}{n-1}}\right)\bigg]\,,
\end{align}
where $c$ is a constant given by
\begin{equation}
    c=\frac{n-1}{V_0^{1/(n-1)}}\left(\frac{2}{n^2}\right)^{n/2(n-1)}\,.
\end{equation}

It is known that the general k-essence Lagrangian admits scaling solutions only if it can be written as
\be
\label{eq:scaling_k-essence}
    P(\varphi,\tilde{X})=\tilde{X}g(\tilde{Y})\,,
\ee
which is satisfied for
\begin{equation}
\label{eq:g_powerlaw}
    g(\tilde{Y})=-\frac{U_0}{\tilde{Y}}+c\,\tilde{Y}^{(2-n)/2(n-1)}\,,
\end{equation}
provided the following two conditions hold:
\ba
    k&=&\frac{2n}{n-2}\,,\\
    U&=&U_0e^{-k\varphi}\,.
\ea
Using Eq.~\eqref{eq:varphi}, the potential corresponds to
\be
\label{eq:U}
U(\phi)=U_0\phi^{2n/(n-2)}\,,
\ee
which is exactly the power law \eqref{eq:U_power_law} obtained in the hybrid three form formulation. Substituting it in Eq.~\eqref{eq:L0} leads to the same Lagrangian~\eqref{eq:pure_scalar}, hence demonstrating the consistency between the three-form and k-essence frameworks in describing scaling cosmologies.

\subsection{Scaling density parameter $\Omega_\varphi$}

To further the comparison between the two frameworks, we now derive, as an example, the density parameter $\Omega_\varphi$ of the k-essence scalar field and verify that it coincides with $\Omega_A$, the one obtained in the three-form dynamical system analysis. 

Along a scaling solution of the k-essence Lagrangian $P(\varphi,\tilde{X}$) in Eq.~\eqref{eq:scaling_k-essence}, the density parameter $\Omega_\varphi$ is given by~\cite{Tsujikawa:2006mw},
\begin{equation}
    \Omega_\varphi=3\gamma\frac{g(\tilde{Y})+\tilde{Y}g'(\tilde{Y})}{k^2}\,,
\end{equation}
where $g^\prime=dg/d\tilde{Y}$. Using Eq.~\eqref{eq:g_powerlaw}, we find
\begin{equation}
\label{eq:Omega_varphi}
    \Omega_\varphi=\frac{3}{8n}\frac{(n-2)^2}{V_0^{1/(n-1)}}\left(\frac{2}{n^2}\right)^{n/2(n-1)}\tilde{Y}^{(2-n)/2(n-1)}\,,
\end{equation}
where $\tilde{Y}$ remains constant during the scaling regime.

The density parameter can be re-expressed back in terms of the hybrid potentials, $U(\chi)$ and $V(\phi)$. Using Eqs.~\eqref{eq:V}, \eqref{eq:varphi}, \eqref{eq:X_tilde}, \eqref{eq:Y}, and \eqref{eq:U}, we find
\begin{equation}
\label{eq:Y_bis}
    \tilde{Y} = \frac{n^2}{2}V_0^{2/n}U_0^{2(n-1)/n}\left(\frac{U}{V}\right)^{2(1-n)/n}\,.
\end{equation}
Substituting this expression in Eq.~\eqref{eq:Omega_varphi}, we obtain the density parameter, valid for scaling with either radiation or non-relativistic matter:
\begin{equation}
\label{eq:Omega_scaling2}
    \Omega_\varphi=3V_0^{-2/n}U_0^{(2-n)/n}\frac{(n-2)^2}{4n^3}\gamma\left(\frac{U}{V}\right)^{(n-2)/n}\,.
\end{equation}

The ratio $U/V$ can be determined from the existence condition of the scaling fixed point in k-essence~\cite{Tsujikawa:2006mw},
\begin{equation}
    (2-\gamma)g(\tilde{Y})-2(\gamma-1)\tilde{Y}g'(\tilde{Y})=0\,,
\end{equation}
where $g$ and $\tilde{Y}$ are given by Eqs.~\eqref{eq:g_powerlaw} and \eqref{eq:Y_bis}, respectively. We find
\begin{equation}
    \frac{U}{V}=\frac{n}{\gamma}-1\,,
\end{equation}
for $n>\gamma$, which reproduces Eq.~\eqref{eq:UoverV1}. Inserting this result into Eq.~\eqref{eq:Omega_scaling2} yields,
\begin{equation}
\Omega_\varphi =\frac{3M^2}{k^2}\frac{\gamma}{n}\left(\frac{n}{\gamma}-1\right)^{2/k}\,,
\end{equation}
noting, as in Eq.~\eqref{eq:M},
\begin{equation}
    M=V_0^{-1/n}U_0^{(2-n)/2n}\,.
\end{equation}
Therefore ${\Omega_\varphi=\Omega_A}$ is the scaling density parameter in Eq.~\eqref{eq:Omega_scaling1} that we found within the hybrid framework in Sec. \ref{sec:fixed-points}. This outcome confirms that the scaling regimes in k-essence and three-form formulations are cosmologically equivalent.

\bibliography{bib}

@article{Tsujikawa:2004dp,
    author = "Tsujikawa, Shinji and Sami, M.",
    title = "{A Unified approach to scaling solutions in a general cosmological background}",
    eprint = "hep-th/0409212",
    archivePrefix = "arXiv",
    doi = "10.1016/j.physletb.2004.10.023",
    journal = "Phys. Lett. B",
    volume = "603",
    pages = "113--123",
    year = "2004"
}

@article{Piazza:2004df,
    author = "Piazza, Federico and Tsujikawa, Shinji",
    title = "{Dilatonic ghost condensate as dark energy}",
    eprint = "hep-th/0405054",
    archivePrefix = "arXiv",
    reportNumber = "BICOCCA-FT-04-4",
    doi = "10.1088/1475-7516/2004/07/004",
    journal = "JCAP",
    volume = "07",
    pages = "004",
    year = "2004"
}

@article{Tsujikawa:2006mw,
    author = "Tsujikawa, Shinji",
    title = "{General analytic formulae for attractor solutions of scalar-field dark energy models and their multi-field generalizations}",
    eprint = "hep-th/0601178",
    archivePrefix = "arXiv",
    doi = "10.1103/PhysRevD.73.103504",
    journal = "Phys. Rev. D",
    volume = "73",
    pages = "103504",
    year = "2006"
}

@article{Nunes:2003ff,
    author = "Nunes, Nelson J. and Lidsey, James E.",
    title = "{Reconstructing the dark energy equation of state with varying alpha}",
    eprint = "astro-ph/0310882",
    archivePrefix = "arXiv",
    doi = "10.1103/PhysRevD.69.123511",
    journal = "Phys. Rev. D",
    volume = "69",
    pages = "123511",
    year = "2004"
}

@article{Copeland:1997et,
    author = "Copeland, Edmund J. and Liddle, Andrew R and Wands, David",
    title = "{Exponential potentials and cosmological scaling solutions}",
    eprint = "gr-qc/9711068",
    archivePrefix = "arXiv",
    reportNumber = "SUSX-TH-97-022, SUSSEX-AST-97-11-1, PU-RCG-97-20",
    doi = "10.1103/PhysRevD.57.4686",
    journal = "Phys. Rev. D",
    volume = "57",
    pages = "4686--4690",
    year = "1998"
}

@article{Wetterich:1987fm,
    author = "Wetterich, C.",
    title = "{Cosmology and the Fate of Dilatation Symmetry}",
    eprint = "1711.03844",
    archivePrefix = "arXiv",
    primaryClass = "hep-th",
    reportNumber = "PRINT-87-0756, DESY-87-123",
    doi = "10.1016/0550-3213(88)90193-9",
    journal = "Nucl. Phys. B",
    volume = "302",
    pages = "668--696",
    year = "1988"
}

@inproceedings{Wands:1993zm,
    author = "Wands, David and Copeland, Edmund J. and Liddle, Andrew R.",
    title = "{Exponential potentials, scaling solutions and inflation}",
    booktitle = "{16th Texas Symposium on Relativistic Astrophysics and 3rd Particles, Strings, and Cosmology Symposium}",
    reportNumber = "SUSSEX-AST-93-1-2",
    pages = "0647--652",
    month = "3",
    year = "1993"
}

@article{Koivisto:2009fb,
    author = "Koivisto, Tomi S. and Nunes, Nelson J.",
    title = "{Inflation and dark energy from three-forms}",
    eprint = "0908.0920",
    archivePrefix = "arXiv",
    primaryClass = "astro-ph.CO",
    doi = "10.1103/PhysRevD.80.103509",
    journal = "Phys. Rev. D",
    volume = "80",
    pages = "103509",
    year = "2009"
}

@article{Koivisto:2009ew,
    author = "Koivisto, Tomi S. and Nunes, Nelson J.",
    title = "{Three-form cosmology}",
    eprint = "0907.3883",
    archivePrefix = "arXiv",
    primaryClass = "astro-ph.CO",
    doi = "10.1016/j.physletb.2010.01.051",
    journal = "Phys. Lett. B",
    volume = "685",
    pages = "105--109",
    year = "2010"
}

@article{Wetterich:1994bg,
    author = "Wetterich, Christof",
    title = "{The Cosmon model for an asymptotically vanishing time dependent cosmological 'constant'}",
    eprint = "hep-th/9408025",
    archivePrefix = "arXiv",
    reportNumber = "HD-THEP-94-16",
    journal = "Astron. Astrophys.",
    volume = "301",
    pages = "321--328",
    year = "1995"
}

@article{Zlatev:1998tr,
    author = "Zlatev, Ivaylo and Wang, Li-Min and Steinhardt, Paul J.",
    title = "{Quintessence, cosmic coincidence, and the cosmological constant}",
    eprint = "astro-ph/9807002",
    archivePrefix = "arXiv",
    doi = "10.1103/PhysRevLett.82.896",
    journal = "Phys. Rev. Lett.",
    volume = "82",
    pages = "896--899",
    year = "1999"
}

@article{Chiba:1999ka,
    author = "Chiba, Takeshi and Okabe, Takahiro and Yamaguchi, Masahide",
    title = "{Kinetically driven quintessence}",
    eprint = "astro-ph/9912463",
    archivePrefix = "arXiv",
    reportNumber = "UTAP-352",
    doi = "10.1103/PhysRevD.62.023511",
    journal = "Phys. Rev. D",
    volume = "62",
    pages = "023511",
    year = "2000"
}

@article{dePutter:2007ny,
    author = "de Putter, Roland and Linder, Eric V.",
    title = "{Kinetic k-essence and Quintessence}",
    eprint = "0705.0400",
    archivePrefix = "arXiv",
    primaryClass = "astro-ph",
    doi = "10.1016/j.astropartphys.2007.05.011",
    journal = "Astropart. Phys.",
    volume = "28",
    pages = "263--272",
    year = "2007"
}

@article{SupernovaCosmologyProject:1998vns,
    author = "Perlmutter, S. and others",
    collaboration = "Supernova Cosmology Project",
    title = "{Measurements of $\Omega$ and $\Lambda$ from 42 High Redshift Supernovae}",
    eprint = "astro-ph/9812133",
    archivePrefix = "arXiv",
    reportNumber = "LBNL-41801, LBL-41801",
    doi = "10.1086/307221",
    journal = "Astrophys. J.",
    volume = "517",
    pages = "565--586",
    year = "1999"
}

@article{SupernovaSearchTeam:1998fmf,
    author = "Riess, Adam G. and others",
    collaboration = "Supernova Search Team",
    title = "{Observational evidence from supernovae for an accelerating universe and a cosmological constant}",
    eprint = "astro-ph/9805201",
    archivePrefix = "arXiv",
    doi = "10.1086/300499",
    journal = "Astron. J.",
    volume = "116",
    pages = "1009--1038",
    year = "1998"
}

@article{Caldwell:1997ii,
    author = "Caldwell, R. R. and Dave, Rahul and Steinhardt, Paul J.",
    title = "{Cosmological imprint of an energy component with general equation of state}",
    eprint = "astro-ph/9708069",
    archivePrefix = "arXiv",
    doi = "10.1103/PhysRevLett.80.1582",
    journal = "Phys. Rev. Lett.",
    volume = "80",
    pages = "1582--1585",
    year = "1998"
}

@article{Copeland:2006wr,
    author = "Copeland, Edmund J. and Sami, M. and Tsujikawa, Shinji",
    title = "{Dynamics of dark energy}",
    eprint = "hep-th/0603057",
    archivePrefix = "arXiv",
    doi = "10.1142/S021827180600942X",
    journal = "Int. J. Mod. Phys. D",
    volume = "15",
    pages = "1753--1936",
    year = "2006"
}

@article{Steinhardt:1999nw,
    author = "Steinhardt, Paul J. and Wang, Li-Min and Zlatev, Ivaylo",
    title = "{Cosmological tracking solutions}",
    eprint = "astro-ph/9812313",
    archivePrefix = "arXiv",
    doi = "10.1103/PhysRevD.59.123504",
    journal = "Phys. Rev. D",
    volume = "59",
    pages = "123504",
    year = "1999"
}

@article{Peebles:2002gy,
    author = "Peebles, P. J. E. and Ratra, Bharat",
    editor = "Hsu, Jong-Ping and Fine, D.",
    title = "{The Cosmological Constant and Dark Energy}",
    eprint = "astro-ph/0207347",
    archivePrefix = "arXiv",
    reportNumber = "KSUPT-02-3",
    doi = "10.1103/RevModPhys.75.559",
    journal = "Rev. Mod. Phys.",
    volume = "75",
    pages = "559--606",
    year = "2003"
}

@article{Barreiro:1999zs,
    author = "Barreiro, T. and Copeland, Edmund J. and Nunes, N. J.",
    title = "{Quintessence arising from exponential potentials}",
    eprint = "astro-ph/9910214",
    archivePrefix = "arXiv",
    reportNumber = "SUSX-TH-016",
    doi = "10.1103/PhysRevD.61.127301",
    journal = "Phys. Rev. D",
    volume = "61",
    pages = "127301",
    year = "2000"
}

@article{Morais:2016bev,
    author = "Morais, Jo\~ao and Bouhmadi-L\'opez, Mariam and Sravan Kumar, K. and Marto, Jo\~ao and Tavakoli, Yaser",
    title = "{Interacting 3-form dark energy models: distinguishing interactions and avoiding the Little Sibling of the Big Rip}",
    eprint = "1608.01679",
    archivePrefix = "arXiv",
    primaryClass = "gr-qc",
    doi = "10.1016/j.dark.2016.11.002",
    journal = "Phys. Dark Univ.",
    volume = "15",
    pages = "7--30",
    year = "2017"
}

@article{Bouhmadi-Lopez:2016dzw,
    author = "Bouhmadi-L\'opez, Mariam and Marto, Jo\~ao and Morais, Jo\~ao and Silva, C\'esar M.",
    title = "{Cosmic infinity: A dynamical system approach}",
    eprint = "1611.03100",
    archivePrefix = "arXiv",
    primaryClass = "gr-qc",
    doi = "10.1088/1475-7516/2017/03/042",
    journal = "JCAP",
    volume = "03",
    pages = "042",
    year = "2017"
}

@article{BeltranAlmeida:2018nin,
    author = "Beltr\'an Almeida, Juan P. and Guarnizo, Alejandro and Valenzuela-Toledo, C\'esar A.",
    title = "{Arbitrarily coupled $p-$forms in cosmological backgrounds}",
    eprint = "1810.05301",
    archivePrefix = "arXiv",
    primaryClass = "astro-ph.CO",
    reportNumber = "PI/UAN-2018-640FT",
    doi = "10.1088/1361-6382/ab5f3c",
    journal = "Class. Quant. Grav.",
    volume = "37",
    number = "3",
    pages = "035001",
    year = "2020"
}

@article{daFonseca:2024boz,
    author = "da Fonseca, Vitor and Barros, Bruno J. and Barreiro, Tiago and Nunes, Nelson J.",
    title = "{Non-canonical 3-form dark energy}",
    eprint = "2410.11658",
    archivePrefix = "arXiv",
    primaryClass = "gr-qc",
    doi = "10.1016/j.dark.2025.101827",
    journal = "Phys. Dark Univ.",
    volume = "47",
    pages = "101827",
    year = "2025"
}

@article{Duff:1980qv,
    author = "Duff, M. J. and van Nieuwenhuizen, P.",
    title = "{Quantum Inequivalence of Different Field Representations}",
    reportNumber = "ICTP/79-80/33",
    doi = "10.1016/0370-2693(80)90852-7",
    journal = "Phys. Lett. B",
    volume = "94",
    pages = "179--182",
    year = "1980"
}

@article{Wongjun:2016tva,
    author = "Wongjun, Pitayuth",
    title = "{Perfect fluid in Lagrangian formulation due to generalized three-form field}",
    eprint = "1602.00682",
    archivePrefix = "arXiv",
    primaryClass = "gr-qc",
    doi = "10.1103/PhysRevD.96.023516",
    journal = "Phys. Rev. D",
    volume = "96",
    number = "2",
    pages = "023516",
    year = "2017"
}

@article{Wongjun:2017spo,
    author = "Wongjun, Pitayuth",
    title = "{Generalized Three-Form Field}",
    eprint = "1708.05795",
    archivePrefix = "arXiv",
    primaryClass = "gr-qc",
    doi = "10.1088/1742-6596/883/1/012002",
    journal = "J. Phys. Conf. Ser.",
    volume = "883",
    number = "1",
    pages = "012002",
    year = "2017"
}

@article{Germani:2009iq,
    author = "Germani, Cristiano and Kehagias, Alex",
    title = "{P-nflation: generating cosmic Inflation with p-forms}",
    eprint = "0902.3667",
    archivePrefix = "arXiv",
    primaryClass = "astro-ph.CO",
    doi = "10.1088/1475-7516/2009/03/028",
    journal = "JCAP",
    volume = "03",
    pages = "028",
    year = "2009"
}

@ARTICLE{1922ZPhy...10..377F,
       author = {{Friedmann}, A.},
        title = "{{\"U}ber die Kr{\"u}mmung des Raumes}",
      journal = {Zeitschrift fur Physik},
         year = 1922,
        month = jan,
       volume = {10},
        pages = {377-386},
          doi = {10.1007/BF01332580}
}

@ARTICLE{1931MNRAS..91..483L,
       author = {{Lema{\^\i}tre}, G.},
        title = "{A homogeneous universe of constant mass and increasing radius accounting for the radial velocity of extra-galactic nebulae}",
      journal = {mnras},
         year = 1931,
        month = mar,
       volume = {91},
        pages = {483-490},
          doi = {10.1093/mnras/91.5.483}
}

@ARTICLE{1936ApJ....83..257R,
       author = {{Robertson}, H.~P.},
        title = "{Kinematics and World-Structure III.}",
      journal = {apj},
         year = 1936,
        month = may,
       volume = {83},
        pages = {257},
          doi = {10.1086/143726},
       adsurl = {https://ui.adsabs.harvard.edu/abs/1936ApJ....83..257R}
}

@article{SravanKumar:2016biw,
    author = "Sravan Kumar, K. and Mulryne, David J. and Nunes, Nelson J. and Marto, Jo\~ao and Vargas Moniz, Paulo",
    title = "{Non-Gaussianity in multiple three-form field inflation}",
    eprint = "1606.07114",
    archivePrefix = "arXiv",
    primaryClass = "astro-ph.CO",
    doi = "10.1103/PhysRevD.94.103504",
    journal = "Phys. Rev. D",
    volume = "94",
    number = "10",
    pages = "103504",
    year = "2016"
}

@article{Koivisto:2009sd,
    author = "Koivisto, Tomi S. and Mota, David F. and Pitrou, Cyril",
    title = "{Inflation from N-Forms and its stability}",
    eprint = "0903.4158",
    archivePrefix = "arXiv",
    primaryClass = "astro-ph.CO",
    doi = "10.1088/1126-6708/2009/09/092",
    journal = "JHEP",
    volume = "09",
    pages = "092",
    year = "2009"
}

@article{DESI:2024mwx,
    author = "Adame, A. G. and others",
    collaboration = "DESI",
    title = "{DESI 2024 VI: cosmological constraints from the measurements of baryon acoustic oscillations}",
    eprint = "2404.03002",
    archivePrefix = "arXiv",
    primaryClass = "astro-ph.CO",
    reportNumber = "FERMILAB-PUB-24-0154-PPD",
    doi = "10.1088/1475-7516/2025/02/021",
    journal = "JCAP",
    volume = "02",
    pages = "021",
    year = "2025"
}

@article{Cortes:2024lgw,
    author = "Cort{\^e}s, Marina and Liddle, Andrew R.",
    title = "{Interpreting DESI's evidence for evolving dark energy}",
    eprint = "2404.08056",
    archivePrefix = "arXiv",
    primaryClass = "astro-ph.CO",
    doi = "10.1088/1475-7516/2024/12/007",
    journal = "JCAP",
    volume = "12",
    pages = "007",
    year = "2024"
}

@article{Wolf:2024eph,
    author = "Wolf, William J. and Garc{\'\i}a-Garc{\'\i}a, Carlos and Bartlett, Deaglan J. and Ferreira, Pedro G.",
    title = "{Scant evidence for thawing quintessence}",
    eprint = "2408.17318",
    archivePrefix = "arXiv",
    primaryClass = "astro-ph.CO",
    doi = "10.1103/PhysRevD.110.083528",
    journal = "Phys. Rev. D",
    volume = "110",
    number = "8",
    pages = "083528",
    year = "2024"
}

@article{Rubin:2023jdq,
    author = "Rubin, David and others",
    title = "{Union Through UNITY: Cosmology with 2,000 SNe Using a Unified Bayesian Framework}",
    eprint = "2311.12098",
    archivePrefix = "arXiv",
    primaryClass = "astro-ph.CO",
    doi = "10.3847/1538-4357/adc0a5",
    journal = "Astrophys. J.",
    volume = "986",
    number = "2",
    pages = "231",
    year = "2025"
}

@article{Scolnic:2021amr,
    author = "Scolnic, Dan and others",
    title = "{The Pantheon+ Analysis: The Full Data Set and Light-curve Release}",
    eprint = "2112.03863",
    archivePrefix = "arXiv",
    primaryClass = "astro-ph.CO",
    doi = "10.3847/1538-4357/ac8b7a",
    journal = "Astrophys. J.",
    volume = "938",
    number = "2",
    pages = "113",
    year = "2022"
}

@article{DES:2024jxu,
    author = "Abbott, T. M. C. and others",
    collaboration = "DES",
    title = "{The Dark Energy Survey: Cosmology Results with {\ensuremath{\sim}}1500 New High-redshift Type Ia Supernovae Using the Full 5 yr Data Set}",
    eprint = "2401.02929",
    archivePrefix = "arXiv",
    primaryClass = "astro-ph.CO",
    reportNumber = "FERMILAB-PUB-23-0821-PPD, DES-2023-805",
    doi = "10.3847/2041-8213/ad6f9f",
    journal = "Astrophys. J. Lett.",
    volume = "973",
    number = "1",
    pages = "L14",
    year = "2024"
}

@article{Planck:2018vyg,
    author = "Aghanim, N. and others",
    collaboration = "Planck",
    title = "{Planck 2018 results. VI. Cosmological parameters}",
    eprint = "1807.06209",
    archivePrefix = "arXiv",
    primaryClass = "astro-ph.CO",
    doi = "10.1051/0004-6361/201833910",
    journal = "Astron. Astrophys.",
    volume = "641",
    pages = "A6",
    year = "2020",
    note = "[Erratum: Astron.Astrophys. 652, C4 (2021)]"
}

@article{ACT:2023dou,
    author = "Qu, Frank J. and others",
    collaboration = "ACT",
    title = "{The Atacama Cosmology Telescope: A Measurement of the DR6 CMB Lensing Power Spectrum and Its Implications for Structure Growth}",
    eprint = "2304.05202",
    archivePrefix = "arXiv",
    primaryClass = "astro-ph.CO",
    reportNumber = "FERMILAB-PUB-23-237-PPD, FERMILAB-PUB-23-237-PPD",
    doi = "10.3847/1538-4357/acfe06",
    journal = "Astrophys. J.",
    volume = "962",
    number = "2",
    pages = "112",
    year = "2024"
}

@article{ACT:2023kun,
    author = "Madhavacheril, Mathew S. and others",
    collaboration = "ACT",
    title = "{The Atacama Cosmology Telescope: DR6 Gravitational Lensing Map and Cosmological Parameters}",
    eprint = "2304.05203",
    archivePrefix = "arXiv",
    primaryClass = "astro-ph.CO",
    reportNumber = "FERMILAB-PUB-23-206-PPD",
    doi = "10.3847/1538-4357/acff5f",
    journal = "Astrophys. J.",
    volume = "962",
    number = "2",
    pages = "113",
    year = "2024"
}

@article{Linder:2025zxb,
    author = "Linder, Eric V.",
    title = "{Uplifting, Depressing, and Tilting Dark Energy}",
    eprint = "2506.02122",
    archivePrefix = "arXiv",
    primaryClass = "astro-ph.CO",
    journal = "",
    month = "6",
    year = "2025"
}

@article{daFonseca:2023ury,
    author = "da Fonseca, Vitor and Barreiro, Tiago and Nunes, Nelson J.",
    title = "{Relaxing cosmological constraints on current neutrino masses}",
    eprint = "2311.01803",
    archivePrefix = "arXiv",
    primaryClass = "astro-ph.CO",
    doi = "10.1103/PhysRevD.109.063517",
    journal = "Phys. Rev. D",
    volume = "109",
    number = "6",
    pages = "063517",
    year = "2024"
}

@article{daFonseca:2021imp,
    author = "da Fonseca, Vitor and Barreiro, Tiago and Nunes, Nelson J.",
    title = "{A simple parametrisation for coupled dark energy}",
    eprint = "2104.14889",
    archivePrefix = "arXiv",
    primaryClass = "astro-ph.CO",
    doi = "10.1016/j.dark.2021.100940",
    journal = "Phys. Dark Univ.",
    volume = "35",
    pages = "100940",
    year = "2022"
}

@article{Barros:2022kpo,
    author = "Barros, Bruno J. and da Fonseca, Vitor",
    title = "{Coupling quintessence kinetics to electromagnetism}",
    eprint = "2209.12189",
    archivePrefix = "arXiv",
    primaryClass = "astro-ph.CO",
    doi = "10.1088/1475-7516/2023/06/048",
    journal = "JCAP",
    volume = "06",
    pages = "048",
    year = "2023"
}

@article{daFonseca:2022qdf,
    author = "da Fonseca, Vitor and others",
    title = "{Fundamental physics with ESPRESSO: Constraining a simple parametrisation for varying $\alpha$}",
    eprint = "2204.02930",
    archivePrefix = "arXiv",
    primaryClass = "astro-ph.CO",
    doi = "10.1051/0004-6361/202243795",
    journal = "Astron. Astrophys.",
    volume = "666",
    pages = "A57",
    year = "2022"
}

@article{CosmoVerseNetwork:2025alb,
    author = "Di Valentino, Eleonora and others",
    collaboration = "CosmoVerse Network",
    title = "{The CosmoVerse White Paper: Addressing observational tensions in cosmology with systematics and fundamental physics}",
    eprint = "2504.01669",
    archivePrefix = "arXiv",
    primaryClass = "astro-ph.CO",
    doi = "10.1016/j.dark.2025.101965",
    journal = "Phys. Dark Univ.",
    volume = "49",
    pages = "101965",
    year = "2025"
}

@article{Barros:2023nzr,
    author = "Barros, Bruno J. and Beltr{\'a}n Jim{\'e}nez, Jose",
    title = "{Non-trivial thick brane realisations with 3-forms}",
    eprint = "2312.12516",
    archivePrefix = "arXiv",
    primaryClass = "gr-qc",
    doi = "10.1007/JHEP02(2024)002",
    journal = "JHEP",
    volume = "02",
    pages = "002",
    year = "2024"
}

@misc{Bouhmadi-Lopez:2025lzm,
    author = "Bouhmadi-L{\'o}pez, Mariam and Chiang, Hsu-Wen and Boiza, Carlos G. and Chen, Pisin",
    title = "{Observational constraints on 3-forms dark energy}",
    eprint = "2512.09991",
    archivePrefix = "arXiv",
    primaryClass = "astro-ph.CO",
    month = "12",
    year = "2025"
}

@article{Bouhmadi-Lopez:2020wve,
    author = "Bouhmadi-L\'opez, Mariam and Chen, Che-Yu and Chew, Xiao Yan and Ong, Yen Chin and Yeom, Dong-Han",
    title = "{Regular Black Hole Interior Spacetime Supported by Three-Form Field}",
    eprint = "2005.13260",
    archivePrefix = "arXiv",
    primaryClass = "gr-qc",
    doi = "10.1140/epjc/s10052-021-09080-1",
    journal = "Eur. Phys. J. C",
    volume = "81",
    number = "4",
    pages = "278",
    year = "2021"
}

@article{Bouhmadi-Lopez:2021zwt,
    author = "Bouhmadi-L\'opez, Mariam and Chen, Che-Yu and Chew, Xiao Yan and Ong, Yen Chin and Yeom, Dong-han",
    title = "{Traversable wormhole in Einstein 3-form theory with self-interacting potential}",
    eprint = "2108.07302",
    archivePrefix = "arXiv",
    primaryClass = "gr-qc",
    doi = "10.1088/1475-7516/2021/10/059",
    journal = "JCAP",
    volume = "10",
    pages = "059",
    year = "2021"
}

@article{Bouhmadi-Lopez:2018lly,
    author = "Bouhmadi-L\'opez, Mariam and Brizuela, David and Garay, I\~naki",
    title = "{Quantum behavior of the ''Little Sibling'' of the Big Rip induced by a three-form field}",
    eprint = "1802.05164",
    archivePrefix = "arXiv",
    primaryClass = "gr-qc",
    doi = "10.1088/1475-7516/2018/09/031",
    journal = "JCAP",
    volume = "09",
    pages = "031",
    year = "2018"
}

@article{Morais:2017vlf,
    author = "Morais, Jo\~ao and Bouhmadi-L\'opez, Mariam and Marto, Jo\~ao",
    editor = {D\k{a}browski, Mariusz P. and Kr\"amer, Manuel and Salzano, Vincenzo},
    title = "{3-Form Cosmology: Phantom Behaviour, Singularities and Interactions}",
    doi = "10.3390/universe3010021",
    journal = "Universe",
    volume = "3",
    number = "1",
    pages = "21",
    year = "2017"
}

\end{document}